%% file: ms.tex
\documentclass[]{emulateapj}
\input{Definitions.tex}

\usepackage{amsmath}
\begin{document}

\title{IMPLODING IGNITION WAVES. I. ONE-DIMENSIONAL ANALYSIS}

\author{Doron Kushnir\altaffilmark{1}, Eli Livne\altaffilmark{2},
and Eli Waxman\altaffilmark{1}} \altaffiltext{1}{Department of Particle Physics and Astrophysics, Weizmann Institute of Science, Rehovot 76100, Israel} \altaffiltext{2}{Racah Institute of Physics, Hebrew University, Jerusalem, Israel}

\begin{abstract}

We show that converging spherical and cylindrical shock waves may ignite a detonation wave in a combustible medium, provided the radius at which the shocks become strong exceeds a critical radius, $R_{\textrm{crit}}$. An approximate analytic expression for $R_{\textrm{crit}}$ is derived for an ideal gas equation of state and a simple (power-law-Arrhenius) reaction law, and shown to reproduce the results of numerical solutions. For typical acetylene--air experiments we find $R_{\textrm{crit}}\sim100\,\mu\textrm{m}$ (spherical) and $R_{\textrm{crit}}\sim1\,\textrm{mm}$ (cylindrical). We suggest that the deflagration to detonation transition (DDT) observed in these systems may be due to converging shocks produced by the turbulent deflagration flow, which reaches sub (but near) sonic velocities on scales $\gg R_{\textrm{crit}}$. Our suggested mechanism differs from that proposed by Zel'dovich et al., in which a fine-tuned spatial gradient in the chemical induction time is required to be maintained within the turbulent deflagration flow. Our analysis may be readily extended to more complicated equations of state and reaction laws. An order of magnitude estimate of $R_{\textrm{crit}}$ within a white dwarf at the pre-detonation conditions believed to lead to Type Ia supernova explosions is 0.1 km, suggesting that our proposed mechanism may be relevant for DDT initiation in these systems. The relevance of our proposed ignition mechanism to DDT initiation may be tested by both experiments and numerical simulations.

\end{abstract}


\keywords{hydrodynamics -- shock waves -- supernovae: individual (Ia)}


\section{Introduction}
\label{sec:Introduction}

The nature of the physical mechanisms driving a deflagration to detonation transition (DDT) is a basic open question in combustion theory \citep[see][for reviews]{Lee1977igd,lee1978mtd,williams1985combustion,lewis1987cfe,shepherd1992tdd,kuo2005poc}. DDT in the laboratory has been observed in several types of experiments in chambers containing cold, unreacted, exothermic gases. These include igniting the mixture with a spark \citep[with or without obstacles along the walls of the channels, see, e.g.,][]{oppenheim1962onset,laderman1963ninth,urtiew1966eot,peraldi1988criteria,teodorczyk1995fast,dorofeev1996deflagration,dorofeev2002flame} or creating a series of shock--flame interactions \citep[e.g.,][]{markstein1964nonsteady,scarinci1990some,scarinci1993amp,thomas1997experimental,thomas2001eof}. Experiments attempting to observe unconfined DDT \citep[e.g.,][]{knystautas1978seventeenth} showed that a transition to detonation induced by turbulent flames in systems without walls or obstacles is rather difficult.

The most extreme case where a DDT may occur is in a Type Ia supernova (SNIa) explosion, which is completely unconfined \citep[see][ for a review]{hillebrandt2000typeIa}. One-dimensional detonation models of SNIa \citep{arnett1969pms,hansen1969cwd} do not reproduce the observed spectra. Both one-dimensional deflagration models \citep{nomoto1976cds,nomoto1984awd,woosley1986pse} and delayed-detonation models, in which a DDT occurs
(\citealp{woosley1990ddm}\footnote{Similar to a model previously proposed by A.~M.~ Khokhlov.} \citealp{khokhlov1991dtm,yamaoka1992ldm,khokhlov1993lct,arnett1994ddm,arnett1994ddmb,woosley1994mss,hoflich1995ddm,hoeflich1996emt,iwamoto1999ncm})
are claimed to fit the observational data.

DDTs observed in the laboratory are commonly characterized by turbulent flow fields, initiated by the interaction of the laminar flame with walls, obstacles, or shocks, and by the initiation of a detonation wave by a sudden explosion (``hot spot''). Two-dimensional \citep{livne1993nsc,arnett1994ddm,khokhlov1995ptf,niemeyer1995tnf,niemeyer1996ocd,reinecke1999tec} and three-dimensional \citep{reinecke2002rnm,reinecke2002tds,gamezo2003tss,gamezo2004ddt,gamezo2005tdd} SNIa flame simulations show that the gravity-induced Rayleigh--Taylor instability makes the flame turbulent, which allows extensive interpenetration of burnt and unburnt materials. The formation of hot spots in this case was not demonstrated computationally, presumably because of the small scales of the hot spots, which are not resolved in the simulations.

The hot spots observed in the laboratory are believed to be produced by a mechanism related to that proposed by \citet{zeldovich1970ood}, in which a spontaneous reaction wave propagates through a reactive material with a spatial gradient in chemical induction time that leads to a deflagration propagating at the Chapman--Jouguet (CJ) speed \citep[see also][]{lee1978photochemical}. However, it is not clear how the required finely tuned spatial gradient is obtained. This ``preconditioning'' of the unburnt material appears to be artificial. Moreover, for both the laboratory experiments and SNIa explosions, it is not clear how such a gradient can be maintained fixed in the turbulent deflagration field over the critical length required for a successful ignition \citep[see, however,][]{oran2007origins}.

Current studies of DDT use direct numerical simulations to investigate the possibility of the initiation of a detonation in turbulent deflagration flames. Due to the challenging computational requirements, calculations are typically performed under simplifying approximations, and the initiation of detonation is typically not directly demonstrated. Rather, the turbulent flow field properties are studied and phenomenological arguments are used to argue for the plausibility of the initiation of detonation \citep[e.g.,][]{woosley2009tis,poludnenko2011stt,woosley2011fts}.

In this paper, we consider the ignition of a detonation wave by converging spherical and cylindrical shock waves, imploding onto an unburnt combustible medium. We limit our analysis to simple (ideal gas) equation of state and reaction laws, which allows us to investigate the various ignition modes and the flow behavior over a wide range of values of the relevant parameters. The model is described in detail in Section~\ref{sec:model}. The model equations are given in Section~\ref{sec:model_eqs}, dimensional analysis is used in Section~\ref{sec:dimensional_analysis} to determine the characteristic length and timescales of the problem, and a summary of the dimensionless parameters is given in Section~\ref{sec:dimensionless_eqs}. The conditions under which a successful ignition is obtained are derived analytically in Section~\ref{sec:estimates}, and investigated numerically in Section~\ref{sec:numerical} (the details of the numerical model are given in Appendix~\ref{sec:numerical model}). We find that the radius of the shock at fixed shock velocity should exceed a critical radius in order to achieve a successful ignition.

We note that ignition of detonation by converging shocks has been considered before, in studies of ignition by shock focusing through reflection off concave walls \citep[e.g.,][]{borisov1990igc,chan1990idi,bartenev2000ert,gelfand2000ddi} and by converging toroidal shocks in "pulse detonation engines" \citep{li2003dip,jackson2003wii,jackson2005gdi,li2005dia}. However, these studies were mostly experimental, while analytical and numerical investigations were limited to the study of the large-scale behavior of the flow in the experimental setup. In particular, a possible connection between the converging flows and the generation of much smaller scale hot spots that may trigger detonation, as well as the hot spot properties required for detonation ignition, were not considered.

The analysis in this paper is one dimensional (limited to spherical or cylindrical symmetry). In a subsequent paper currently in preparation, we show that the evolution of multidimensional perturbations during shock implosion does not suppress the ignition of a detonation wave. Although our analysis is limited to simple equations of state and reaction laws, our predictions for the critical radius may be tested experimentally, since the properties of many explosive gases are well described by our simplified model. Moreover, our analysis may be readily extended to more complicated equations of state and reaction laws. Such extension for the case of pre-detonations in SNIa explosions, taking into consideration the effects of the large differences between the reaction lengths of the different burning stages in that case (see Section~\ref{sec:Discussion}), is presented in a third paper (in preparation).

Our results are summarized and their possible implications to DDTs are discussed in Section~\ref{sec:Discussion}.


\section{The model}
\label{sec:model}

In this section, we describe the model in detail. The model equations are given in Section~\ref{sec:model_eqs}, dimensional analysis is used in Section~\ref{sec:dimensional_analysis} to determine the characteristic length and timescales of the problem, and a summary of the dimensionless parameters is given in Section~\ref{sec:dimensionless_eqs}.

\subsection{Model Equations}
\label{sec:model_eqs}

The reactive Euler equations, describing one-dimensional flow of an ideal gas that undergoes a single irreversible reaction, are
\begin{eqnarray}
\label{eq:hydro_eq}
(\partial_{t}+u\partial_{r})\ln\rho+ r^{-(\nu-1)}\partial_{r}(r^{\nu-1}u) &=& 0,
\nonumber \\
(\partial_{t}+u\partial_{r})u+\rho^{-1}\partial_{r}p &=&
0, \nonumber \\
(\partial_{t}+u\partial_{r})\left(\frac{p}{\rho(\gamma-1)}\right) +\frac{p}{\rho r^{\nu-1}}\partial_{r}(r^{\nu-1}u)&=& QW, \nonumber \\
(\partial_{t}+u\partial_{r})\lambda &=& W,
\end{eqnarray}
where $\nu=1,2,3$ for planar, cylindrical, and spherical symmetry, respectively. The total internal energy of the gas is given by
\begin{eqnarray}\label{eq:eos}
\varepsilon(p,V,\lambda)=\frac{p}{\rho(\gamma-1)}-\lambda Q,
\end{eqnarray}
where $p$ is the pressure, $\rho$ is the density, $u$ is the fluid velocity, $\lambda$ is the fraction of burnt material, $\gamma$ is the ratio of specific heats, and $Q$ is the heat release per unit mass. We assume a reaction rate of the form
\begin{eqnarray}\label{eq:rate}
W=\kappa\left(\frac{\rho}{\rho_{\star}}\right)^{n}(1-\lambda)^{m} e^{-\frac{\rho}{\rho_{\star}}\frac{p_{A}}{p}},
\end{eqnarray}
where $\rho_{\star}$, $\kappa$, $n$, $m$, and $p_{A}$ are parameters characterizing the reaction. The argument of the exponential in Equation~\eqref{eq:rate} is the usual Arrhenius law for the temperature, $-T_{A}/T$, where $T_{A}$ is the activation temperature, written in terms of density and pressure.

In the planer case ($\nu=1$), the reactive Euler equations have solutions, the so called ZND waves \citep[][]{zel1940theory,von1942theory,doring1943theory}, where a strong shock ignites the fuel and the burning proceeds to equilibrium in a reaction zone behind the shock, while the energy released continues to drive the shock. In the case where the flow at the end of the reaction zone is sonic, the detonation is called a CJ detonation, and the detonation propagates at the CJ velocity, given by $D_{\textrm{CJ}}=\sqrt{2Q(\gamma^{2}-1)}$. We define the dimensionless parameter
\begin{eqnarray}\label{eq:tau def}
\tau=\frac{p_{A}}{\rho_{0}D_{\textrm{CJ}}^{2}}=\frac{p_{A}}{(\gamma+1)p_{\textrm{CJ}}},
\end{eqnarray}
where $p_{\textrm{CJ}}=\rho_{0}D_{\textrm{CJ}}^{2}/(\gamma+1)$ is the CJ pressure (achieved at the end of the reaction zone), and $\rho_{0}$ is the initial density of the gas (without loss of generality, we may assume $\rho_{\star}=\rho_{0}$, where difference in their values is absorbed into $\kappa$ and into $p_{A}$). $\tau$ measures how gradually the fuel burns in a detonation wave, as explained below. Let us define $x_{0.5(0.9)}$ as the distance behind the shock front for which 0.5 (0.9) of the fuel is burned (the reaction zone length is $x_{1}$). Then, for small values of $\tau$ the fuel burns slowly behind the shock ($x_{0.9}/x_{0.5}\rightarrow\infty$), while as $\tau\rightarrow\infty$ the profile of the burnt fraction behind the shock tends to a step function, in which a reactionless induction zone is terminated by a very thin ``fire'' in which all chemical energy is released ($x_{0.9}/x_{0.5}\rightarrow1$).

The initial conditions consist of a strong (spherical or cylindrical) shock wave, imploding from infinity, onto an unburnt stationary fluid with an initial density $\rho_{0}$.

\subsection{Characteristic Length and Timescales}
\label{sec:dimensional_analysis}

Let us first consider the pure hydrodynamic case (i.e., $Q=0$). In this case, the equations contain no characteristic length scales, and as the shock converges the flow approaches Guderley's self-similar (SlS) solution \citep{Guderley42}. This solution describes the flow for both the imploding stage (before the shock reaches the center) and the exploding stage (after the shock is reflected from the center as a finite Mach number shock). In what follows we analyze these two stages, as both are relevant for the ignition of a detonation wave. We denote the imploding (exploding) stage with a subscript $i$ ($e$).

In the SlS solution, the shock position, $R$, is given by the time relative to the time at which it reaches the center, $t_{0}$, as a power law
\begin{eqnarray}\label{eq:R(t)}
R\propto\left(t-t_{0}\right)^{\alpha}.
\end{eqnarray}
Equivalently, the shock velocity is related to the shock position through $\dot{R}\propto R^{\delta}$, where $\delta=(\alpha-1)/\alpha$. The value of $\alpha$ is determined by requiring the flow in the implosion stage to include a sonic point at some position $r_{s}(t)=\zeta_{s}R(t)$. The resulting value of $\alpha$ is smaller than one ($\delta$ is negative), such that the shock accelerates to an infinite velocity as it converges to the center. Some values of $\alpha$, $\delta$, and $\zeta_{s}$ as functions of $\gamma$ and $\nu$ are given in Table~\ref{tbl:alpha_delta_values}. In what follows, we define $R_{\textrm{CJ}}$ as the radius at which $\dot{R}=-D_{\textrm{CJ}}$ during the implosion, and define $\xi\equiv R/R_{\textrm{CJ}}$,
\begin{eqnarray}\label{eq:RCJ}
\dot{R}=A_{j}D_{\textrm{CJ}}\left(\frac{R}{R_{\textrm{CJ}}}\right)^{\delta}\equiv A_{j}D_{\textrm{CJ}}\xi^{\delta},
\end{eqnarray}
where $j=i$ or $j=e$, $A_{i}=-1$, and $0<A_{e}<1$ (some values of $A_{e}$ are given in Table~\ref{tbl:alpha_delta_values}). At $R_{\textrm{CJ}}$ the internal energy generated by the hydrodynamic shock is comparable to $Q$ (see Equation~\eqref{eq:Qh}).

The SlS solution and the Rankine--Hugoniot relations determine the post-shock density, pressure, and (the laboratory frame) velocity to be
\begin{eqnarray}\label{eq:RH}
\rho_{s,j}&=&f_{\rho,j}\frac{\gamma+1}{\gamma-1}\rho_{0},\nonumber\\
p_{s,j}&=&f_{p,j}\frac{2}{\gamma+1}\rho_{0}\dot{R}^{2},\nonumber\\
u_{s,j}&=&f_{u,j}\frac{2}{\gamma+1}\dot{R},
\end{eqnarray}
where $\dot{R}$ is the shock velocity, $f_{\rho,i}=f_{p,i}=f_{u,i}=1$, $f_{\rho,e}>1$, $f_{p,e}>1$, and $0<f_{u,e}<1$ (some values of $f_{\rho,e}$, $f_{p,e}$, and $f_{u,e}$ are given in Table~\ref{tbl:alpha_delta_values}).

\begin{deluxetable}{cccccc}
\tablecaption{Some Values of $\alpha$, $\delta$, $\xi_{0}$, $\zeta_{s}$, $A_{e}$, $f_{\rho,e}$, $f_{p,e}$, and $f_{u,e}$ as Functions of $\gamma$ and $\nu$ \label{tbl:alpha_delta_values}}
\tablewidth{0pt} \tablehead{ \colhead{$\gamma$} & \colhead{} & \colhead{$1.25$} & \colhead{$4/3$}  & \colhead{$5/3$}}
\startdata $\nu=2$ & $\alpha$  & 0.85  & 0.84  & 0.82\\
                   & $\delta$  & --0.17 & --0.19 & --0.23\\
                   & $\xi_{0}$ & 0.10  & 0.22  & 1.00\\
                   & $\zeta_{s}$ & 1.15 & 1.16  & 1.20\\
                   & $A_{e}$ & 0.21 & 0.29  & 0.58\\
                   & $f_{\rho,e}$ & 32.5 & 18.0  & 5.7\\
                   & $f_{p,e}$ & 822 & 272  & 29.5\\
                   & $f_{u,e}$ & 0.17 & 0.18  & 0.21\\
           $\nu=3$ & $\alpha$ & 0.74  & 0.73  & 0.69\\
                   & $\delta$ & --0.34 & --0.37 & --0.45\\
                   & $\xi_{0}$ & 0.31  & 0.47 & 1.00\\
                   & $\zeta_{s}$ & 1.11 & 1.12  & 1.15\\
                   & $A_{e}$ & 0.21 & 0.30  & 0.64\\
                   & $f_{\rho,e}$ & 91 & 38  & 8.1\\
                   & $f_{p,e}$ & $1.65\times10^{3}$ & 411  & 30.6\\
                   & $f_{u,e}$ & 0.22 & 0.24  & 0.29\\
\enddata
\end{deluxetable}

We turn now to analyzing the $Q>0$ case. Let us introduce a hydrodynamical time scale, $t_{h,j}$, and an energy per unit mass scale, $Q_{h,j}$, characterizing the pure hydrodynamic ($Q=0$) flow, and a timescale characterizing the chemical energy release, $t_{q,j}$. As long as
\begin{eqnarray}\label{eq:hydro condition}
\frac{Q}{t_{q,j}}\min(t_{q,j},t_{h,j})\ll Q_{h,j},
\end{eqnarray}
the chemical energy generation term in the third equation of Equations~\eqref{eq:hydro_eq} may be neglected, such that Equations~\eqref{eq:hydro_eq} reduce to the pure hydrodynamic Euler equations, i.e., describing the pure hydrodynamic case. We assume that initially $t_{h,i}\ll t_{q,i}$ and that Equation~\eqref{eq:hydro condition} holds through the entire flow, such that the flow approaches the SlS solution. We define $Q_{h,j}$ as the post-shock internal energy excluding chemical energy,
\begin{eqnarray}\label{eq:Qh}
Q_{h,j}&\equiv&\frac{p}{\rho(\gamma-1)}\simeq\frac{f_{p,j}}{f_{\rho,j}}\frac{2\dot{R}^{2}}{(\gamma+1)^{2}}\nonumber\\
&=&\frac{2f_{p,j}}{f_{\rho,j}}\left(\frac{A_{j}D_{\textrm{CJ}}}{\gamma+1}\right)^{2} \xi^{2\delta}.
\end{eqnarray}
We further define
\begin{eqnarray}\label{eq:th}
t_{h,j}\equiv f_{h,j}\frac{\alpha R}{\dot{R}}
=\frac{f_{h,j}\alpha}{A_{j}} \frac{R_{\textrm{CJ}}}{D_{\textrm{CJ}}}\xi^{1-\delta},
\end{eqnarray}
where the dimensionless $f_{h,j}$ is determined as follows. At the implosion stage, the available hydrodynamical time for a fluid element to release chemical energy sufficient for affecting the flow is the crossing time between the shock front and the sonic point (the flow behind the sonic point cannot affect the flow in front of it), which can be estimated as $\simeq-(\zeta_{s}-1)R/u_{s,i}=-(\gamma+1)(\zeta_{s}-1)R/2\dot{R}$. This estimate leads to the definition
\begin{equation}\label{eq:f_hi}
    f_{h,i}=-(\gamma+1)(\zeta_{s}-1)/2\alpha.
\end{equation}
At the explosion stage the entire flow behind the shock is subsonic, and the hydrodynamical timescale may be estimated simply as the time passed from the shock's reflection, i.e., $f_{h,e}=1$.

The post-shock chemical energy release timescale is
\begin{eqnarray}\label{eq:tq}
t_{q,j}&\equiv&\left(\frac{1}{\lambda}\frac{d\lambda}{dt}\right)^{-1}
\simeq\frac{1}{\kappa} \left(\frac{\rho}{\rho_{0}}\right)^{-n}e^{\frac{\rho}{\rho_{0}}
\frac{p_{A}}{p}}\nonumber\\
&\simeq&\frac{1}{\kappa}\left(f_{\rho,j}\frac{\gamma+1}{\gamma-1}\right)^{-n}e^{\eta \frac{f_{\rho,j}}{f_{p,j}}A_{j}^{-2}\xi^{-2\delta}},
\end{eqnarray}
where
\begin{equation}\label{eq:eta}
    \eta\equiv\tau(\gamma+1)^{2}/2(\gamma-1).
\end{equation}
If the inequality of Equation~\eqref{eq:hydro condition} is violated during the implosion, deviations from the pure hydrodynamical solution, e.g., ignition of a detonation wave, may arise. If  Equation~\eqref{eq:hydro condition} holds throughout the implosion, no significant deviation from the hydrodynamic solution arises, and a detonation wave is not ignited. In this case, the inequality of Equation~\eqref{eq:hydro condition} holds at the onset of explosion and may be violated as the shock explodes and $Q_{h}$ decreases.

\subsection{Dimensionless Parameters}
\label{sec:dimensionless_eqs}

The dimensional parameters defining the problem are $\rho_{0}$, $R_{\textrm{CJ}}$, $D_{\textrm{CJ}}$, $\kappa$, and $p_{A}$. Based on dimensional arguments, these five-dimensional parameters may be used to construct three independent dimensional parameters (using which a dimensionless parameter may not be constructed). Hence, the solutions are completely determined (up to scaling of the measurement units of length, time, and mass) by a set of six dimensionless parameters: $\nu$, $n$, $m$, and $\gamma$, which appear in the equations, and two additional parameters constructed from the five-dimensional parameters. We choose the latter two to be $\tau$ (or $\eta$, which is linear in $\tau$, see Equation~\eqref{eq:eta}) and
\begin{eqnarray}\label{eq:theta}
\theta&\equiv&\ln\left(\frac{\kappa\alpha R_{\textrm{CJ}}}{D_{\textrm{CJ}}}\right)+n\ln\left(\frac{\gamma+1}{\gamma-1}\right)+2\delta\ln \xi_{0},
\end{eqnarray}
where
\begin{eqnarray}\label{eq:xi0}
\xi_{0}&\equiv&\left[\frac{4(\gamma-1)}{\gamma+1}\right]^{-1/2\delta}.
\end{eqnarray}
Some values of $\xi_{0}$ as a function of $\gamma$ and $\nu$ are given in Table~\ref{tbl:alpha_delta_values}.

The motivation for the choice of $\theta$, which is essentially the ratio of $R_{\textrm{CJ}}$ to $D_{\textrm{CJ}}/\kappa$, is clarified in Section~\ref{sec:estimates}. We show in particular that ignition is obtained for values of $\theta$ exceeding a minimum, or ``critical'', value $\theta_{c}$, and derive $\theta_{c}$ as a function of the other five dimensionless parameters. The critical value of $\theta$, $\theta_{c}$, corresponds, using Equation~\eqref{eq:theta}, to a critical value of $R_{\textrm{CJ}}$. That is, a successful ignition requires $R_{\textrm{CJ}}$ (the radius at which the shock velocity reaches $D_{\textrm{CJ}}$) to exceed a critical value. Using Equation~(\ref{eq:RCJ}), the lower limit for $R_{\textrm{CJ}}$ may be expressed as a lower limit for the radius at which the shock reaches any chosen velocity.


\section{Simple analytic estimates}
\label{sec:estimates}

In this section, we present an approximate analytic analysis of the conditions under which a successful ignition is obtained. We  define the ignition criterion, show that it implies that ignition is obtained for values of $\theta$ exceeding a minimum value $\theta_{c}$, and derive $\theta_{c}$ as a function of the dimensionless parameters. A detailed analysis of the flow behavior for different values of $\theta$ is given in Appendices~\ref{sec:implosion_stage} and~\ref{sec:explosion_stage} for the implosion and explosion stages, respectively. In particular, the radii at which ignition is expected to be achieved are estimated.

We define our successful ignition criterion as achieving chemical energy release that exceeds the hydrodynamical thermal energy generation at some shock radius. This requirement may be written as
\begin{eqnarray}\label{eq:hydro condition2}
\frac{Q}{t_{q,j}}\min(t_{q,j},f_{t,j}t_{h,j})>f_{Q,j} Q_{h,j},
\end{eqnarray}
where $f_{t,j}$ and $f_{Q,j}$ are order unity functions of the dimensionless parameters (to be calibrated from numerical simulations, Section~\ref{sec:numerical}), and the hydrodynamic time, $t_{h,j}$, the combustion time, $t_{q,j}$, and the hydrodynamic energy release, $Q_{h,j}$, were defined in Section~\ref{sec:dimensional_analysis}. Note that $t_{h,j}\ll t_{q,j}$ for both $\xi\rightarrow0$ and $\xi\rightarrow\infty$ (see Equations~\eqref{eq:th} and~\eqref{eq:tq}), and that the condition $t_{q,j}<t_{h,j}$, which may be satisfied within some $\xi$ range, is neither necessary nor sufficient for satisfying the ignition criterion of Equation~\eqref{eq:hydro condition2}.

During explosion, the shock may propagate into material which underwent significant combustion, even if ignition was not achieved. For the explosion stage we therefore require the criterion of Equation~\eqref{eq:hydro condition2} to be satisfied at a radius where the reflected shock propagates into unburnt material, i.e., at a radius for which $t_{q}=\lambda/(d\lambda/dt)$ at the upstream of the shock is larger than $t_{h,e}$ (see Section~\ref{sec:explosion_stage}).

The requirement of Equation~(\ref{eq:hydro condition2}) is a necessary requirement, since otherwise the chemical energy release does not affect the flow (Equation~\eqref{eq:hydro condition} holds), but not necessarily a sufficient one. In Section~\ref{sec:numerical}, we demonstrate that the inequality of Equation~\eqref{eq:hydro condition2} is actually a sufficient condition by investigating numerically a few representative cases: $n=0$, $m=1$, $\gamma=5/3$, $\tau$ in the range $[0.05,2]$ for both cylindrical ($\nu=2$) and spherical $(\nu=3)$ geometries (numerical results for $\{n=1, m=1, \gamma=1.25, \tau=0.7, \theta\simeq\theta_{c}, \nu=2,3\}$, values relevant for laboratory experiments, are given in Section~\ref{sec:Discussion}). We also calibrate in Section~\ref{sec:numerical} the values of $f_{t,j}$ and $f_{Q,j}$ for the examined parameters.

As noted above, $t_{h,j}\ll t_{q,j}$ for both $\xi\rightarrow0$ and $\xi\rightarrow\infty$. For $f_{t,j}t_{h,j}<t_{q,j}$ Equation~\eqref{eq:hydro condition2} may be written, using Equations~\eqref{eq:Qh}, ~\eqref{eq:th}, ~\eqref{eq:tq}, ~\eqref{eq:theta}, and~\eqref{eq:xi0}, as
\begin{eqnarray}\label{eq:hydro condition3}
\theta+\Lambda_{j}\equiv\theta+\ln\left(\frac{f_{t,j}f_{h,j}f_{\rho,j}^{n+1}}{f_{Q,j}A_{j}^{3}f_{p,j}}\right)
&>&\nonumber\\
\eta\frac{f_{\rho,j}}{f_{p,j}} A_{j}^{-2}\xi^{-2\delta}-(1-3\delta)\ln \xi&\equiv& g_{j}(\xi).
\end{eqnarray}
The relation between $t_{q,j}$ and $t_{h,j}$ is given by
\begin{eqnarray}\label{eq:tq=th}
\ln\left(\frac{t_{q,j}}{f_{t,j}t_{h,j}}\right)=g_{j}(\xi)-\Lambda_{j}-2\delta\ln\left(\frac{\xi}{\bar{\xi}_{0,j}}\right)-\theta.
\end{eqnarray}
For $t_{q,j}<f_{t,j}t_{h,j}$, which may be obtained during implosion/explosion, Equation~\eqref{eq:hydro condition2} may be written, using Equations~\eqref{eq:Qh} and~\eqref{eq:xi0}, as
\begin{eqnarray}\label{eq:hydro condition3a}
\xi>\bar{\xi}_{0,j}\equiv\left(f_{Q,j} A_{j}^{2}\frac{f_{p,j}}{f_{\rho,j}}\right)^{-1/2\delta}\xi_{0}.
\end{eqnarray}
Thus, Equation~\eqref{eq:hydro condition2} is satisfied if (and only if) Equation~\eqref{eq:hydro condition3} holds for some $\xi>\bar{\xi}_{0,j}$.

\begin{figure}
\epsscale{1} \plotone{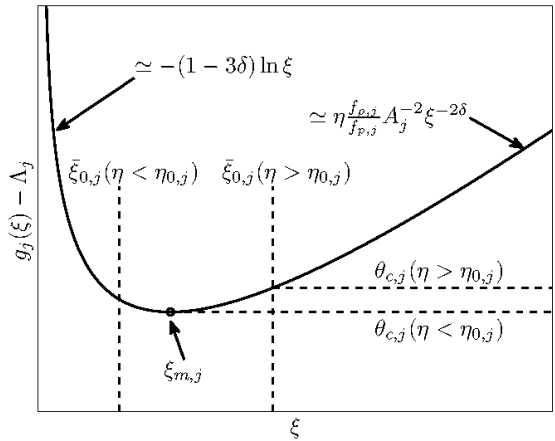} \caption{Qualitative illustration of the function $g_{j}(\xi)$, and of the significance of $\xi_{m,j}$, $\bar{\xi}_{0,j}$, and $\theta_{c,j}$  (see Section~\ref{sec:estimates}).
\label{fig:gxi}}
\end{figure}

An illustration of the function $g_{j}(\xi)$ is shown in Figure~\ref{fig:gxi}. $g_{j}(\xi)$ diverges for both $\xi\rightarrow0$ (as $\simeq-(1-3\delta)\ln \xi$) and $\xi\rightarrow\infty$ (as $\simeq\eta f_{\rho,j}f_{p,j}^{-1} A_{j}^{-2}\xi^{-2\delta}$), and its minimum is obtained at
\begin{eqnarray}\label{eq:xim}
\xi_{m,j}=\left(\frac{1-3\delta}{-2\delta\eta}\right)^{-1/2\delta}\left(\frac{f_{p,j}}{f_{\rho,j}}A_{j}^2\right)^{-1/2\delta}.
\end{eqnarray}
Defining $\xi_{c,j}=\max\{\xi_{m,j},\bar{\xi}_{0,j}\}$, the inequality of Equation~\eqref{eq:hydro condition2} may be written as a lower limit for $\theta$ (see Figure~\ref{fig:gxi}),
\begin{eqnarray}\label{eq:thetac}
\theta>\theta_{c,j}\equiv g_{j}(\xi_{c,j})-\Lambda_{j}.
\end{eqnarray}
Defining
\begin{eqnarray}\label{eq:eta0}
\eta_{0,j}\equiv\frac{1-3\delta}{-2\delta}\xi_0^{2\delta}f_{Q,j}^{-1},
\end{eqnarray}
$\xi_{c,j}$ is given by
\begin{equation}\label{eq:xic}
\xi_{c,j}=\left\{
  \begin{array}{ll}
    \bar{\xi}_{0,j}, & \hbox{$\eta>\eta_{0,j}$} \\
    \xi_{m,j}, & \hbox{$\eta\le\eta_{0,j}$}
  \end{array}
\right.
\end{equation}

The inequality of Equation~(\ref{eq:thetac}) is equivalent to that of Equation~(\ref{eq:hydro condition2}). For the implosion case, this implies that ignition is obtained provided $\theta$ exceeds the critical value $\theta_{c,i}$. For the explosion case, a more stringent requirement for $\theta$ may be obtained since, as noted above, Equation~(\ref{eq:hydro condition2}) is required to be satisfied at $\xi$ for which $t_{q}$ at the upstream of the shock is larger than $t_{h,e}$. As explained in detail in Section~\ref{sec:explosion_stage}, although this requirement may affect significantly the radius at which ignition is achieved, it does not modify the critical value of $\theta$ required for ignition, $\theta_{c,e}$. In what follows, we therefore adopt Equation~(\ref{eq:thetac}) as the successful ignition criterion.

\begin{figure}
\epsscale{1} \plotone{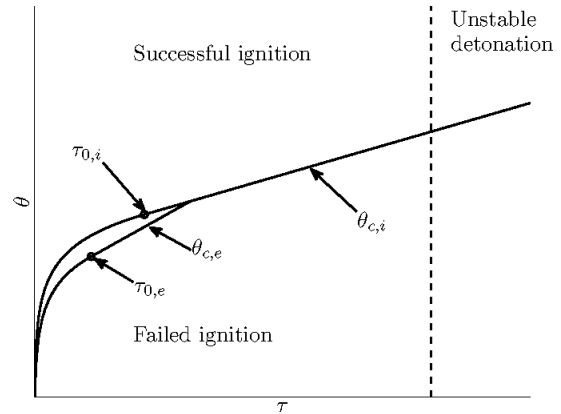} \caption{Qualitative illustration of the critical curves $\theta_{c,j}(\tau)$, Equation~\eqref{eq:thetac}. $\tau$ measures how gradually the fuel burns (see Equation~\eqref{eq:tau def}), and $\theta$ essentially measures the ratio of $R_{\textrm{CJ}}$, the radius at which the shock velocity reaches $D_{\textrm{CJ}}$ (see Equation~\eqref{eq:RCJ}), to $D_{\textrm{CJ}}/\kappa$ (see Equation~\eqref{eq:theta}). $\eta_{0,j}$, Equation~\eqref{eq:eta0}, is related to $\tau_{0,j}$ by $\eta_{0,j}\equiv\tau_{0,j}(\gamma+1)^{2}/2(\gamma-1)$.
\label{fig:estimates}}
\end{figure}

For $\theta_{c,i}<\theta_{c,e}$ ($\theta_{c,i}>\theta_{c,e}$), the critical ignition is obtained in the implosion (explosion) stage (\textit{ignition by implosion (explosion)}) and $\theta_{c}=\theta_{c,i}$ ($\theta_{c}=\theta_{c,e}$). Note that for high enough $\theta$ the ignition is always obtained in the implosion stage. Thus, if for some set of parameters the critical ignition is by explosion, then by increasing $\theta$ a transition to ignition by implosion is obtained for $\theta=\theta_{c,i}$. An qualitative illustration of the critical curves is given in Figure~\ref{fig:estimates}.

For large values of $\tau$ ($\eta$), we have $\bar{\xi}_{0}>\xi_m$ which implies that for $\theta\ge\theta_c$ the flow is significantly affected by chemical energy release at $\xi_d\ge \xi_c=\bar{\xi}_{0}$, where $\theta=g(\xi_{d})-\Lambda$. We expect ignition to take place near $\xi=\xi_d$. Equality, $\xi_d=\xi_c=\bar{\xi}_{0}$, is obtained at critical conditions, i.e., $\theta=\theta_c$. Thus, for large values of $\tau$ and $\theta=\theta_c$, ignition takes place near $R=\bar{\xi}_{0}R_{\textrm{CJ}}$ with $t_h\sim t_q$ and $Q_h\sim Q$. For small values of $\tau$, critical ignition takes place near $\xi={\xi}_m\propto\tau^{-1/2|\delta|}$, which implies that for small $\tau$ ignition is expected at $R\gg R_{\textrm{CJ}}$ with $Q_h/Q\sim\tau\ll1$. For more details see Section~\ref{sec:ig_radii}.

The analysis becomes more complicated for high values of $\tau$, as the detonation wave becomes unstable to one-dimensional and multidimensional perturbations \citep[see][and references therein]{Sharpe1997lsi}. It is not clear that this instability inhibits successful ignitions,
since it can lead to some other stable configuration \citep[i.e., multidimensional detonation cell structure; see][for a review]{fickett2001detonation}. In order to bypass the question of stability in this one-dimensional analysis, we assume that if a detonation wave formed and has propagated over a sufficient distance then the instability will lead to a stable configuration. Since the cell size of cellular detonations is ten to one hundred times the reaction zone length \citep[for both laboratory experiments and SNIa explosions; see, e.g.,][]{lee1984dpg,gamezo1999msc}, we assume that a ``sufficient distance'' is one hundred times of the reaction zone length. This assumption should be examined in a multidimensional analysis and is beyond the scope of this work. However, our results do not depend strongly on the value adopted for the ``sufficient distance''.

Let us summarize briefly the estimates presented in this section. For a choice of $\nu$, $n$, $m$, and $\gamma$, there exists a critical curve in the $\left(\tau,\theta\right)$-plane, given by Equation~\eqref{eq:thetac}. For $\theta>\theta_{c}$ we expect a successful ignition, while for $\theta<\theta_{c}$ we expect a failed ignition. If the critical ignition is by explosion, then by increasing $\theta$ a transition to ignition by implosion is obtained for $\theta=\theta_{c,i}$. For high values of $\tau$, the detonation wave becomes unstable, but we still expect that a detonation wave may propagate for a sufficient distance, provided $\theta>\theta_{c}$. In the example illustrated in Figure~\ref{fig:estimates} the critical curve includes both ignition by implosion and ignition be explosion parts. In general, depending on the values of the dimensionless parameters defining the problem and of $f_{Q,j}$, $f_{t,j}$, the critical curve may completely determined by a single ignition mode. The critical value $\theta_{c}$ implies, equivalently, a critical radius for $R_{\textrm{CJ}}$ (see Equation~\eqref{eq:theta}). Hence, for a successful ignition, there exists a minimal value for $R_{\textrm{CJ}}$, which in turn implies a minimal value for the radius of the imploding shock wave at a given velocity.

\section{Numerical simulations}
\label{sec:numerical}

In this section, we use numerical simulations to study the behavior of the flow at  different parts of the $\left(\tau,\theta\right)$-plane (the details of the numerical model are given in Appendix~\ref{sec:numerical model}). We compare the numerical results with the simple estimates of Section~\ref{sec:estimates}, which also allows us to calibrate the values of $f_{t,j}$ and $f_{Q,j}$. The numerical setup includes a spherical (cylindrical) piston that moves inward into a cold ideal gas. The flow equations are solved numerically, with resolution sufficient for fully resolving the reaction zone. Each simulation is defined by the set of six dimensionless parameters $\nu$, $n$, $m$, $\gamma$, $\tau$, and $\theta$. We focus on the $\nu=3$, $n=0$, $m=1$, and $\gamma=5/3$ case, and investigate the behavior of the flow in four different regions of the $\left(\tau,\theta\right)$-plane. We first demonstrate a successful ignition by implosion in Section~\ref{sec:success}. We then demonstrate in Section~\ref{sec:other} other behaviors of the flow in the $\left(\tau,\theta\right)$-plane: failed ignition due to low $\theta$, a successful ignition by explosion, and a failed ignition due to instability of the detonation wave (high $\tau$). We explain the analysis of the $\left(\tau,\theta\right)$-plane and summarize our results in Section~\ref{sec:numerical analysis}. We show that the analytic estimate of the $\theta_{c,i}$ critical ignition curve, Equation~\eqref{eq:thetac}, provides a good approximation to the numerical curve (better than $0.4$ units in $\theta$ for $0.05\le\tau\le2$) for $f_{t,i}\simeq1.7$ and $f_{Q,i}\simeq0.88$. Similarly, the analytic estimate of $\theta_{c,e}$ provides a good approximation to the numerical curve (up to $0.6$ units in $\theta$ for $0.05\le\tau\le0.8$) for $f_{t,e}\simeq0.41$ and $f_{Q,e}\simeq1.7$. The four cases discussed in Sections~\ref{sec:success}--~\ref{sec:other} are marked in Figure~\ref{fig:Planesa}. We also show the results for the $\nu=2$ case (keeping all other dimensionless parameters the same) in Figure~\ref{fig:Planesb}.

In the remainder of this section, we define $t=0$ as the time at which the shock reaches the center in Guderley's solution. We further normalize the time to $t_{\textrm{CJ}}$, which is minus the time at which the imploding shock reaches $R_{\textrm{CJ}}$.

\subsection{A Successful Ignition by Implosion: $\tau=1$ and $\theta=6.3$}
\label{sec:success}

\begin{figure}
\epsscale{1} \plotone{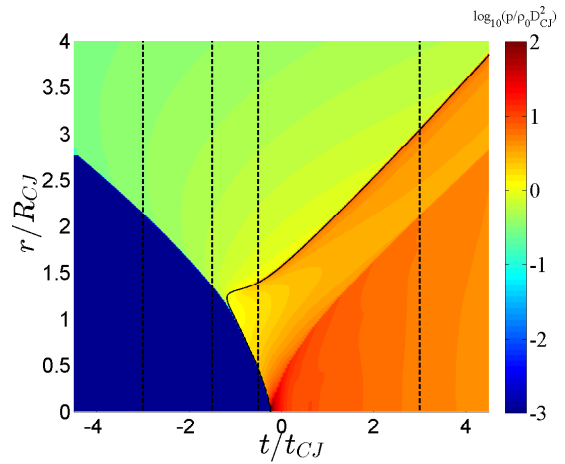} \caption{Normalized pressure map ($\log_{10}(p/\rho_{0}D_{\textrm{CJ}}^{2})$) as a function of $t/t_{\textrm{CJ}}$ and $r/R_{\textrm{CJ}}$, for $\nu=3$, $n=0$, $m=1$, $\gamma=5/3$, $\tau=1$, and $\theta=6.3$. The solid line represents $\lambda=0.5$. The four times shown in Figure~\ref{fig:rig15tau1Prof} are marked with dashed lines.
\label{fig:rig15tau1PrsMapZoom}}
\end{figure}

\begin{figure}
\epsscale{1} \plotone{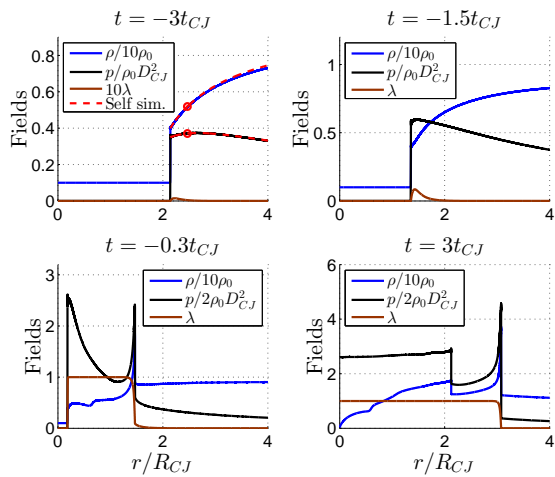} \caption{Normalized density ($\rho/\rho_{0}$, blue), normalized pressure ($p/\rho_{0} D_{\textrm{CJ}}^{2}$, black), and $\lambda$ (brown) as functions of $r/R_{\textrm{CJ}}$ at several times ($t/t_{\textrm{CJ}}=-3,-1.5,-0.3,3$), for $\nu=3$, $n=0$, $m=1$, $\gamma=5/3$, $\tau=1$, and $\theta=6.3$. Guderley's SlS profiles are also shown at $t/t_{\textrm{CJ}}=-3$ (red), along with their singular point (red circle).
\label{fig:rig15tau1Prof}}
\end{figure}

In this section we study a typical case of a successful ignition by implosion, obtained for $\tau=1$ and $\theta=6.3$ (we calibrate $\theta_{c}=\theta_{c,i}=5.85\pm0.05$ for $\tau=1$, see Section~\ref{sec:numerical analysis}). For $f_{t_i}\simeq0.41$ and $f_{Q_i}\simeq0.88$ Equation~\eqref{eq:hydro condition3} is satisfied at the implosion stage at $\xi<\xi_{d}\simeq1.15$ (in this case $\xi_d=\xi_{d,i2}$, see Section~\ref{sec:implosion_stage}). We therefore expect deviations from Guderley's SlS solution as the shock converges below $\xi_{d}$. A normalized pressure ($p/\rho_{0}D_{\textrm{CJ}}^{2}$) map (as a function of $t/t_{\textrm{CJ}}$ and $r/R_{\textrm{CJ}}$), which shows the ignition and the detonation waves, is shown in Figure~\ref{fig:rig15tau1PrsMapZoom}. The solid line in this figure represents $\lambda=0.5$. The spatial profiles of the normalized density ($\rho/\rho_{0}$), normalized pressure ($p/\rho_{0} D_{\textrm{CJ}}^{2}$), and $\lambda$ at several times (marked with dashed lines in Figure~\ref{fig:rig15tau1PrsMapZoom}) are shown in Figure~\ref{fig:rig15tau1Prof}.

At time $t/t_{\textrm{CJ}}=-3$ the converging shock position is $\xi\simeq2.15>\xi_{d}$, and the dynamics are still not affected by the small chemical energy release (see $\lambda$ curves in Figure~\ref{fig:rig15tau1Prof}). As a result, the profiles are consistent with Guderley's SlS profiles (shown in red). Note that since Guderley's SlS profiles include a sonic point (red circle), the region between this point and the piston is supersonic, such that deviations of the numerical solution from the SlS solution are to be expected \citep{WaxmanShvarts93}. Such deviations are seen for the density profiles.

After the shock implodes further to $\xi\simeq1.35\gtrsim\xi_{d}$ ($t/t_{\textrm{CJ}}=-1.5$), the chemical energy release begins to be important, as the burnt fraction increases behind the shock. The numerical profiles begin to show large deviations from Guderley's SlS profiles. Shortly after ($t/t_{\textrm{CJ}}=-0.3$) two detonation waves are formed: one is moving outward (at $r/R_{\textrm{CJ}}\simeq1.45$) into nearly unburnt material (see the small values of $\lambda$ in the upstream of this wave) and the second is moving inward (at $r/R_{\textrm{CJ}}\simeq0.2$). The latter hits the center shortly before $t=0$ (since the detonation wave is faster than the SlS imploding wave) and emerges as a regular shock wave (since the fuel in the upstream for this shock is already burnt). This regular exploding shock is seen at $r/R_{\textrm{CJ}}\simeq2.1$ at $t/t_{\textrm{CJ}}=3$. Note that at this time the pressure tends to a constant value toward the center, while the density tends to zero. This means that the temperature diverges in the center, very similarly to Guderley's SlS solution for the reflected shock. At this time the outgoing detonation wave is already at $r/R_{\textrm{CJ}}\simeq3.1$. This detonation wave meets our criterion for a successful ignition (see Section~\ref{sec:numerical analysis}).

\subsection{Qualitatively Different Behavior of the Flow in Different Parts of the $\left(\tau,\theta\right)$-plane}
\label{sec:other}

\begin{figure}
\epsscale{1} \plotone{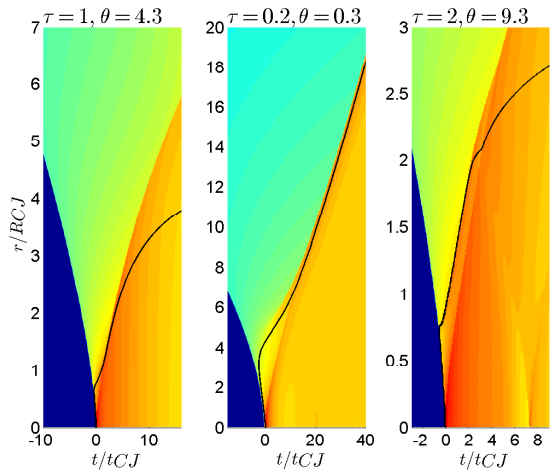} \caption{Normalized pressure maps ($\log_{10}(p/\rho_{0}D_{\textrm{CJ}}^{2})$) as functions of $t/t_{\textrm{CJ}}$ and $r/R_{\textrm{CJ}}$, for $\nu=3$, $n=0$, $m=1$, and $\gamma=5/3$. The color map is the same as in Figure~\ref{fig:rig15tau1PrsMapZoom}, and the solid line represents $\lambda=0.5$. Left panel: $\tau=1$ and $\theta=4.3$. Middle panel: $\tau=0.2$ and $\theta=0.3$. Right panel: $\tau=2$ and $\theta=9.3$.
\label{fig:CasesPrsMap}}
\end{figure}

In this section we investigate the flow in three cases, which are different from the successful ignition by implosion presented in Section~\ref{sec:success}. The first case is a typical case of a failed ignition due to low $\theta$: $\tau=1$ and $\theta=4.3$ (we calibrate $\theta_{c}=\theta_{c,i}=5.85\pm0.05$ for $\tau=1$, see Section~\ref{sec:numerical analysis}). The normalized pressure map for this case is shown in Figure~\ref{fig:CasesPrsMap} (left panel). The material begins to burn when the shock is at $r\simeq R_{\textrm{CJ}}$, but although the burnt fraction increases behind the shock, the chemical energy release cannot affect the hydrodynamics, as it is too small. As a result, no detonation wave is produced, and the converging shock propagates to the center with roughly the same speed as in Guderley's solution (as can be seen in Figure~\ref{fig:CasesPrsMap}, the shock hits the center at $t\simeq0$). After the converging shock hits the center, it emerges as a regular shock wave and later on begins to decelerate. As a result, the pressure decreases, the reaction is suppressed and the ignition fails. In this case our criterion for a successful ignition is not met (see Section~\ref{sec:numerical analysis}).

The second case is a typical case of a successful ignition by explosion with $\tau<\tau_0$: $\tau=0.2$ and $\theta=0.3$ (we calibrate $\theta_{c}=\theta_{c,e}=-0.85\pm0.05$ and $\theta_{c,i}=0.65\pm0.05$ for $\tau=0.2$, see Section~\ref{sec:numerical analysis}). A normalized pressure map for this case is shown in Figure~\ref{fig:CasesPrsMap} (middle panel). We expect ignition to occur during the explosion stage, at a radius $\xi\simeq7$ beyond which significant combustion did not take place (for $f_{Q_e}\simeq1.7$ and $f_{t_e}\simeq1.7$ we obtain $\bar{\xi}_{0,e}\simeq3.0$, $\theta_{0}\simeq-0.71<\theta$ and $\xi_{\textrm{eq},e2}\simeq7.4$, see Section~\ref{sec:explosion_stage}). Indeed, we find numerically that the chemical energy release during the implosion stage is too small for ignition (although the material is burning, see the solid line which represents $\lambda=0.5$), while the reflected shock releases sufficient chemical energy to ignite a successful detonation wave as it moves into unburnt material around $\xi=7$. The resulting detonation wave meets our criterion for a successful ignition (see Section~\ref{sec:numerical analysis}).

The third case is a typical case of a failed ignition due to instability of the detonation wave (high $\tau$): $\tau=2$ and $\theta=9.3$ (we calibrate $\theta_{c}=9.95\pm0.05$ for $\tau=2$, see Section~\ref{sec:numerical analysis}). A normalized pressure map for this case is shown in Figure~\ref{fig:CasesPrsMap} (right panel). In this case although a detonation wave is formed, it fails after some time ($t/t_{\textrm{CJ}}\simeq2.5$). As explained in Section~\ref{sec:estimates}, this failure is attributed to inherent instability of detonation waves with high values of $\tau$. Before the failure of the detonation wave, it manages to propagate only $\simeq45$ reaction zone lengths, such that it does not meet our criterion for a successful ignition (we require propagation over $100$ reaction zone lengths, see Section~\ref{sec:numerical analysis}). Higher values of $\theta$ lead to detonation waves that propagate longer distances, such that they meet our criterion for a successful ignition (see Section~\ref{sec:numerical analysis}).

\subsection{Numerical Analysis of the $\left(\tau,\theta\right)$-plane}
\label{sec:numerical analysis}

In this section, we summarize the analysis of the $\left(\tau,\theta\right)$-plane. Once a criterion for a successful ignition is chosen (see below), the methodology of analysis is as follows. For our numerical analysis, we represent the $\left(\tau,\theta\right)$-plane as a discrete rectangular grid of $\left(\tau,\theta\right)$ points, separated by 0.1 in both the $\tau$- and the $\theta$-directions. For values of $\tau$ in the range of $0.05\le\tau\le2.5$ we searched for the minimal value of $\theta$, $\theta_{s}$, for which a successful ignition is achieved and for the maximal value of $\theta$, $\theta_{f}$, for which ignition is not achieved. The critical value of $\theta$ is then defined as $(\theta_{s}+\theta_{f})/2$. Since our grid resolution is $0.1$, $\theta_c$ is determined to an accuracy of $\pm0.05$, which implies a $\simeq5\%$ accuracy in $R_{\textrm{CJ}}$ (see Equation~\eqref{eq:theta}). If the critical ignition is by explosion (see below for the implosion/explosion criterion) we search for the minimal value of $\theta$ for which the ignition is by implosion ($\theta_{i}$) and the maximal value of $\theta$ for which the ignition is by explosion ($\theta_{e}$). The critical value of $\theta$ for ignition by implosion is then given by $(\theta_{i}+\theta_{e})/2$.

\begin{figure}
\epsscale{1} \plotone{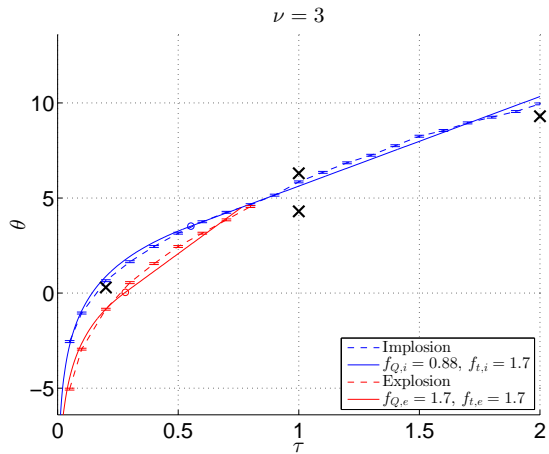} \caption{Numeric (dashed lines) and analytic (full lines, Equation~\eqref{eq:thetac}) critical ignition curves in the $\left(\tau,\theta\right)$-plane for spherical ($\nu=3$) shock, with a reaction law characterized by $\{n=0,m=1\}$ and an ideal gas adiabatic index of $\gamma=5/3$. Blue and red lines show implosion and explosion critical curves, respectively. Error bars reflect the accuracy of our numerical determination of $\theta_{c}$, $\pm0.05$. The analytic curves are obtained from the analysis presented in Section~\ref{sec:estimates}, using $f_{t,i}=1.7$, $f_{Q,i}=0.88$, $f_{t,e}=1.7$, and $f_{Q,e}=1.7$. The circles denote $\tau_{0,j}$. The four cases discussed in Sections~\ref{sec:success}--~\ref{sec:other} are marked with crosses.
\label{fig:Planesa}}
\end{figure}

The derived values of $\theta_{c,i}$ and $\theta_{c,e}$ are shown in Figure~\ref{fig:Planesa} (x's mark the four cases discussed in Sections~\ref{sec:success} and~\ref{sec:other}). The analytic critical curves of $\theta_{c,i}$ and $\theta_{c,e}$, obtained by using the analysis presented in Section~\ref{sec:estimates}, are shown in Figure~\ref{fig:Planesa} for $f_{t,i}=1.7$, $f_{Q,i}=0.88$, $f_{t,e}=1.7$, and $f_{Q,e}=1.7$. These curves provide a good approximation to the numerically inferred $\theta_{c,i}$ and $\theta_{c,e}$ (ignition by implosion: up to $0.4$ units in $\theta$ for $0.05\le\tau\le2$, ignition by explosion: up to $0.6$ units in $\theta$ for $0.05\le\tau\le0.8$).

\begin{figure}
\epsscale{1} \plotone{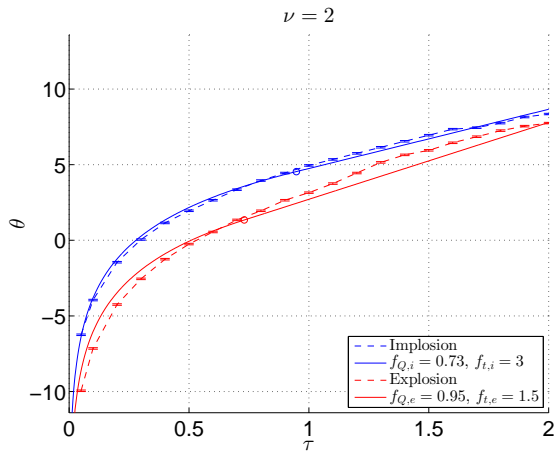} \caption{Same as Figure~\ref{fig:Planesa} for cylindrical ($\nu=2$) shock. The analytic curves are obtained from the analysis presented in Section~\ref{sec:estimates}, using $f_{t,i}=3$, $f_{Q,i}=0.73$, $f_{t,e}=1.5$, and $f_{Q,e}=0.95$.
\label{fig:Planesb}}
\end{figure}

The numeric and the analytic critical curves, $\theta_{c,i}$ and $\theta_{c,e}$, for $\nu=2$ (keeping all other dimensionless parameters the same) are shown in Figure~\ref{fig:Planesb}. The analytic curves obtained for $f_{t,i}=3$, $f_{Q,i}=0.73$, $f_{t,e}=1.5$, and $f_{Q,e}=0.95$ provide a good approximation to the numeric $\theta_{c,i}$ and $\theta_{c,e}$ curves (ignition by implosion: up to $0.3$ units in $\theta$ for $0.05\le\tau\leq2$, ignition by explosion: up to $1.3$ units in $\theta$ for $0.05\le\tau\le2$).

The different functional dependence of $\theta_{c,j}$ on $\tau$ above and below $\tau_{0,j}$, predicted by the analytic analysis (see Equations~\eqref{eq:xic} and~\eqref{eq:thetac}) is well reproduced in the numerical simulations. The $\tau$ dependence of $\xi_{c,j}$, the radius at which ignition is initiated for $\theta=\theta_{c,j}$ (determined from the numerical simulations in a manner similar to that of the determination of $\xi_{d,i2}$ and $\xi_{\textrm{eq},e2}$ in Sections~\ref{sec:success} and~\ref{sec:other}, respectively) is also consistent with the analytic predictions: $\xi_{c,j}$ is roughly given by $\bar{\xi}_{0,j}$ for $\tau>\tau_{0,j}$, and increases consistently with Equation~\eqref{eq:xim} for $\tau<\tau_{0,j}$.

We define a numerical solution in which a successful ignition was achieved as a solution in which a detonation wave has propagated $M_{\textrm{th}}=100$ reaction zone lengths (see Section~\ref{sec:estimates} for the motivation for choosing $M_{\textrm{th}}=100$; our results are not sensitive to the exact value of $M_{\textrm{th}}$, see below). The number of reaction zones propagated was defined as the integral of the distance over which the burning wave propagated as a detonation wave, divided by the instantaneous reaction zone width (defined as the distance between the positions where $\lambda=0.1$ and $\lambda=0.9$ in the simulation). In order to determine whether the burning wave at a given instant is an ``acceptable'' detonation wave, we used the following criterion. We examined the hydrodynamical profiles at times when an outward propagating burning front exists (before it hits the inward moving piston), and measured the distance between the position where $\lambda=0.5$ and the outermost shock wave, $w_{\textrm{Num}}$. We then inspected the upstream density and pressure of the outermost shock wave, and derived the reaction front structure (ZND profile) of a planer CJ detonation wave propagating into a medium with such density and pressure. From the ZND profile we obtained the distance behind the shock front for which $\lambda=0.5$, $w_{\textrm{ZND}}$. Since the detonation waves obtained in our simulations are not exactly at a steady state and are spherical (cylindrical), we expect some difference between $w_{\textrm{Num}}$ and $w_{\textrm{ZND}}$. We chose to define the burning wave as a detonation wave if $w_{\textrm{Num}}/w_{\textrm{ZND}}<w_{\textrm{th}}=3$.

For most of the cases the decision between failure and ignition was not sensitive to the values of $w_{\textrm{th}}$ and $M_{\textrm{th}}$, since in the cases of success we obtained $w_{\textrm{Num}}/w_{\textrm{ZND}}\simeq1$ for many reaction lengths, and in the cases of failure we quickly obtained $w_{\textrm{Num}}/w_{\textrm{ZND}}\gg1$. However, for high values of $\tau$ the derived value of $\theta_{c}$ is somewhat sensitive to the values of $w_{\textrm{th}}$ and $M_{\textrm{th}}$.

In order to decide whether a successful ignition is achieved in a given simulation by implosion or by explosion, we examined the flow at the time for which the detonation wave has propagated $M_{\textrm{th}}$ reaction zone lengths. If a reflected (regular) shock wave exists and it hasn't reached the detonation wave up to this time, we define this case as an ignition by implosion, otherwise as ignition by explosion (typically, the discrimination can be made by inspecting the flow at much earlier times).


\section{Summary and discussion}
\label{sec:Discussion}

We have analyzed the conditions under which converging spherical and cylindrical shock waves ignite a detonation wave in a combustible medium. We have analyzed both analytically and numerically strong shocks imploding onto a combustible medium with an ideal gas equation of state and a simple (power-law-Arrhenius, Equation~\eqref{eq:rate}) reaction law. Given the geometry of the problem ($\nu=2,3$ for cylindrical, spherical symmetry), the adiabatic index $\gamma$ of the gas, and the power-law indices ($\{n,m\}$) determining the reaction rate, the problem is fully defined by two dimensionless parameters: $\tau$ and $\theta$ (see Equations~\eqref{eq:tau def} and~\eqref{eq:theta}). $\tau$ measures how gradually the fuel burns, and $\theta$ essentially measures the ratio of $R_{\textrm{CJ}}$, the radius at which the shock velocity reaches $D_{\textrm{CJ}}$ (see Equation~\eqref{eq:RCJ}), to $D_{\textrm{CJ}}/\kappa$ (see Equation~\eqref{eq:theta}). We have analyzed the different ignition modes (Sections~\ref{sec:estimates} and~\ref{sec:ig_radii}), and have shown that ignition is obtained for values of $\theta$ exceeding a critical value, $\theta_{c}(\tau)$ (Equation~\eqref{eq:thetac}, Figure~\ref{fig:estimates}), and derived approximate analytic expressions for the $\theta_c(\tau)$ curves (Equation~\eqref{eq:thetac}). The analytic expressions provide a good approximation to the exact, numerically derived curves (see Figures~\ref{fig:Planesa} and~\ref{fig:Planesb}).

The minimum value of $\theta$ required for ignition implies a minimum value of $R_{\textrm{CJ}}$ required for ignition. Equivalently, a minimum value of $R_{\textrm{CJ}}$ corresponds to a minimum value of the radius of the shock at any other chosen velocity (see Equation~\eqref{eq:RCJ}). For example, defining the radius at which the shock becomes strong as $R_{2}$, the radius at which the shock velocity is twice the speed of sound $c_{0}$ in the unburned material, we have
\begin{eqnarray}\label{eq:R of M=2}
R_{2}\simeq R_{\textrm{CJ}}\left(\frac{2c_{0}}{D_{\textrm{CJ}}}\right)^{1/\delta}
\end{eqnarray}
\citep[this estimate cannot be used for much weaker shock waves; see][]{ponchaut2006imploding}.

We have shown that ignition may be achieved either during implosion or explosion (following reflection at the center) of the shock wave. The ignition mode at the critical curve depends on the values of $\{n,m,\nu,\gamma\}$, but for high enough values of $\theta$ the ignition is always obtained during implosion. For large values of $\tau$ the detonation wave becomes unstable but nevertheless can propagate for a sufficient distance to guarantee a successful ignition provided $\theta>\theta_{c}$.

Although our analysis is limited to simple equations of state and reaction laws, our predictions for the critical radius may be tested experimentally, since the properties of many explosive gases are well described by our simplified model. Moreover, our analysis may be readily extended to more complicated equations of state and reaction laws (e.g., the case of pre-detonations in SNIa explosions).

This mechanism for the initiation of a detonation wave may be responsible for DDTs. Since near sonic velocities are obtained in the turbulent flow fields typical of DDT on scales much larger than $R_2$ (see below), it is not unreasonable to suggest that the turbulent flow produces converging shocks leading to an ignition of a detonation wave. It is interesting to note that similar ideas regarding ignition by ``hot spots'' have already been proposed by \citet{brinkley1959seventh} and \citet{oppenheim1962onset}, but were abandoned without any quantitative analysis. In order to demonstrate the viability of our proposed DDT mechanism, a detailed analysis of the turbulent flow, which is beyond the scope of this work, is required. The existence of convergent flows around unburnt material needs to be demonstrated, as well as the insensitivity of the ignition to inhomogeneities of the flow fields (density, velocity, burned fraction, etc.).

Let us estimate $R_{\textrm{CJ}}$ and $R_{2}$ for a stoichiometric acetylene--air mix with density of $1.58\times10^{-4}\,\textrm{g}\,\textrm{cm}^{-3}$ at room temperature, as typical for laboratory experiments \citep[see][ for review on relevant experiments, as well as for equation of state and reaction rate prameters]{oran2007origins}. For this gas we have $\gamma=1.25$, $\tau\simeq0.7$, $n=1$, $\kappa\simeq1.58\times10^{8}\,\textrm{s}^{-1}$, and $D_{\textrm{CJ}}\simeq1.87\times10^{5}\,\textrm{cm}\,\textrm{s}^{-1}$. Assuming $f_{t,j}=f_{Q,j}=1$ we find $\tau_{0}\simeq0.66<\tau$ for the spherical case, implying $\xi_{c,i}=\xi_{0}\simeq0.31$ and $\xi_{c,e}=\bar{\xi}_{0,e}\simeq0.21$, from which we obtain $\theta_{c,i}\simeq7.4$ and $\theta_{c,e}\simeq0$. We therefore find that critical ignition is by explosion ($\theta_{c}=\theta_{c,e}$), as should be expected due to the low value of $\gamma-1$, which implies a large amplification of the density by the reflected shock, and to the finite value of $n$, which implies that the reaction rate is sensitive to the density (see Table~\ref{tbl:alpha_delta_values} for the downstream density at the explosion stage). A numerical determination of $\theta_{c}$, obtained by the method described in Section~\ref{sec:numerical}, yields $\theta_{c}=\theta_{c,e}=2.87\pm0.05$, which is in reasonable agreement with our simple estimates. Consequently, we find $R_{\textrm{CJ}}\simeq15\,\mu\textrm{m}$ and, using $c_{0}\simeq3.2\times10^{4}\,\textrm{cm}\,\textrm{s}^{-1}$, $R_{2}\simeq300\,\mu\textrm{m}$. Assuming $f_{t,j}=f_{Q,j}=1$ for the cylindrical case, we have $\tau_{0}\simeq0.98>\tau$, $\xi_{c,i}=\xi_{m,i}\simeq0.25$, $\xi_{c,e}=\xi_{m,e}\simeq0.35$, $\theta_{c,i}\simeq8.1$, and $\theta_{c,e}\simeq1.1$.
Once again, critical ignition is by explosion ($\theta_{c}=\theta_{c,e}$). The numerically determined value, $\theta_{c}=\theta_{c,e}=1.77\pm0.05$, is in a close agreement with our simple estimates, and we find $R_{\textrm{CJ}}\simeq5\,\mu\textrm{m}$ and $R_{2}\simeq2\,\textrm{mm}$.

The estimates given above are consistent with the relevant terrestrial experiments in two senses. First, the scale of $R_{2}$ is much smaller than the size of the experimental channels, typically tens of centimeters, over which near sonic velocities are obtained in the turbulent flow field typical of DDTs \citep[see, e.g.,][]{oran2007origins}. Hence, it is not unreasonable to suggest that the resulting turbulence may produce converging flows on a scale $\sim R_{2}$. Second, $R_{\textrm{CJ}}$ is too small to be fully resolved in three-dimensional simulations \citep[currently reaching cell sizes of only $\simeq100\,\mu\textrm{m}$, see, e.g.,][]{gamezo2005tdr}, which might explain why this ignition mechanism was not observed in simulations (recall that for $\tau>\tau_0$ and ignition by explosion we expect explosion at $R=\bar{\xi}_{0,e} R_{\textrm{CJ}}$).

Similar statements can be made regarding the pre-detonation phase of white-dwarfs in delayed-detonation scenarios of SNIa. For typical pre-detonation conditions, $\rho_{0}=10^{7}\,\textrm{g}\,\textrm{cm}^{-3}$ and equal mass fraction of $^{12}$C and $^{16}$O nuclei, an order of magnitude estimate of $R_2$ (corresponding to a shock velocity $\sim7\times10^{8}\,\textrm{cm}\,\textrm{s}^{-1}$) is 0.1~km, smaller than the typical scale of the nearly sonic turbulence. An order of magnitude estimate of $R_{\textrm{CJ}}$ is 0.01~km, well below the resolution of numerical simulations \citep[which currently reach $\simeq1\,\textrm{km}$ resolution, see, e.g.,][]{gamezo2005tdd}. Note that in this case the detonation wave consists of three burning layers: a Carbon-burning stage, followed by an Oxygen-burning stage and a Silicon-burning stage. The above estimates relate to a successful ignition of a detonation wave with steady carbon and oxygen burning layers. A detailed analysis of the ignition of detonation under conditions relevant for SNIa explosions will be given in a subsequent paper, currently under preparation.

\acknowledgments The authors thank A. M. Khokhlov for useful comments. This research was partially supported by ISF, UPBC, Minerva, and GIF grants.



\appendix


\section{Ignition radii}
\label{sec:ig_radii}

In Section~\ref{sec:estimates}, we presented an approximate analytic analysis of the conditions under
which a successful ignition is obtained. In this section, a detailed analysis of the flow behavior for
different values of $\theta$ is given. In particular, the radii at which ignition is expected to be achieved are
estimated. The implosion and explosion stages are analyzed in Sections~\ref{sec:implosion_stage} and~\ref{sec:explosion_stage}, respectively.

\subsection{The Implosion Stage}
\label{sec:implosion_stage}

\begin{figure}
\epsscale{1} \plottwo{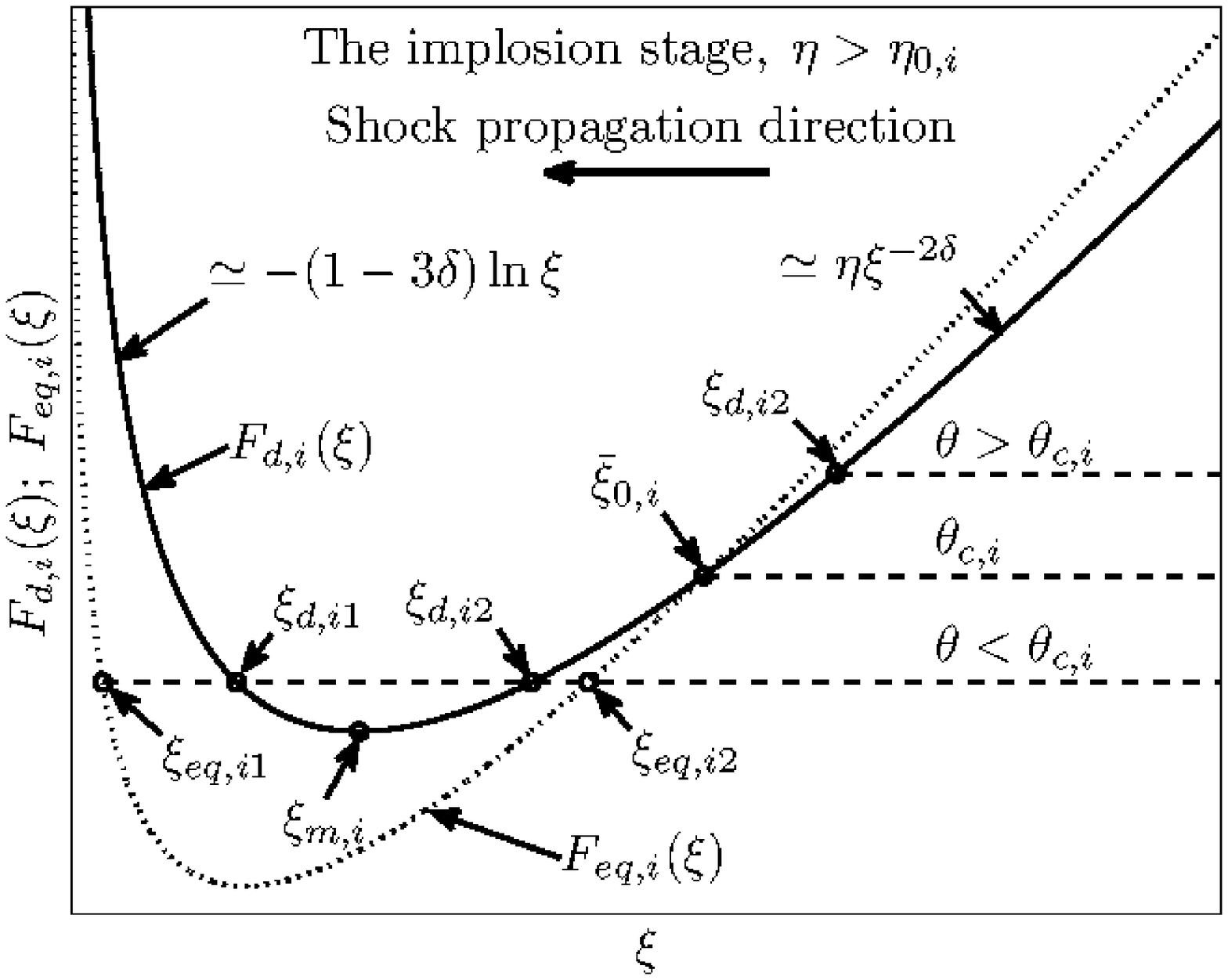}{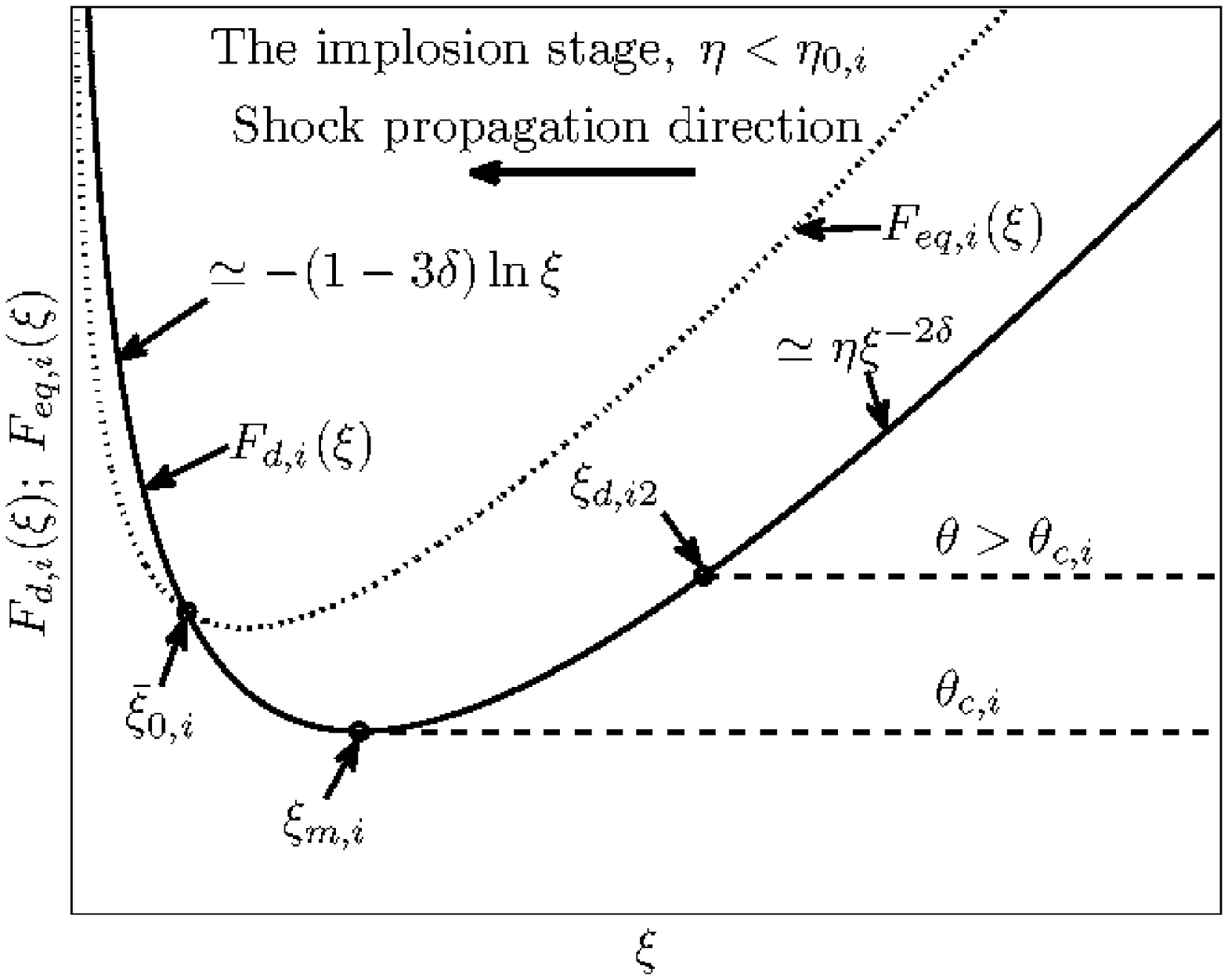} \caption{Qualitative illustration of the functions $F_{d,j}(\xi)=g_{j}(\xi)-\Lambda_{j}$ (solid line) and $F_{\textrm{eq},j}(\xi)=g_{j}(\xi)-\Lambda_{j}-2\delta\ln(\xi/\bar{\xi}_{0,j})$ (dotted line), and of the significance of $\xi_{m,j}$, $\bar{\xi}_{0,j}$, and $\theta_{c,j}$ for the implosion stage with $\eta>\eta_{0,i}$ (left panel) and $\eta<\eta_{0,i}$ (right panel). Dashed lines represent different values of $\theta$. At $\xi=\xi_d$ we have $\theta=F_d$, which implies that the hydrodynamic energy release and the chemical energy release are similar, and at $\xi=\xi_{\textrm{eq}}$ we have $\theta=F_{\textrm{eq}}$, which implies that the hydrodynamic and chemical timescales are similar (see Section~\ref{sec:estimates}).
\label{fig:gxi_imp1}}
\end{figure}

An illustration of the functions $F_{d,i}(\xi)\equiv g_{i}(\xi)-\Lambda_{i}$ and $F_{\textrm{eq},i}(\xi)\equiv g_{i}(\xi)-\Lambda_{i}-2\delta\ln(\xi/\bar{\xi}_{0,i})$ is shown in Figure~\ref{fig:gxi_imp1}. Let us consider first the case $\eta>\eta_{0,i}$. The horizontal lines representing different values of $\theta$ may intersect each of the functions $F_{d,i}(\xi)$ and $F_{\textrm{eq},i}(\xi)$ up to two times. We denote the intersection points (in case they exist) as $\xi_{d,i1}<\xi_{d,i2}$ for $F_{d,i}(\xi)$ and $\xi_{\textrm{eq},i1}<\xi_{\textrm{eq},i2}$ for $F_{\textrm{eq},i}(\xi)$. For $\xi>\xi_{\textrm{eq},i2}$ or $\xi<\xi_{\textrm{eq},i1}$ we have $f_{t,i}t_{h,i}<t_{q,i}$ and Equation~\eqref{eq:hydro condition2} is satisfied for $\theta>F_{d,i}(\xi)$. Since $\eta>\eta_{0,i}$ we have $\bar{\xi}_{0,i}>\xi_{m,i}$ and $\theta_{c,i}= F_{d,i}(\bar{\xi}_{0,i})=F_{\textrm{eq},i}(\bar{\xi}_{0,i})$. For $\theta>\theta_{c,i}$ the first intersection, i.e., the largest $\xi$ intersection, is at $\xi=\xi_{d,i2}$, which implies that ignition is expected at $\xi=\xi_{d,i2}$.

Similar considerations lead to the conclusion that ignition is achieved at $\xi=\xi_{d,i2}$ also for $\eta<\eta_{0,i}$. Thus, for the implosion stage ignition is achieved with $f_{t,i}t_{h,i}\le t_{q,i}$. Equality, $f_{t,i}t_{h,i}= t_{q,i}$, is obtained at $\theta=\theta_c$ and $\eta>\eta_{0,i}$. As $\theta\rightarrow\infty$ we have $\xi_{d,i2}\simeq(\theta/\eta)^{-1/2\delta}\rightarrow\infty$, and ignition is obtained for $R\gg R_{\textrm{CJ}}$.

\begin{figure}
\plottwo{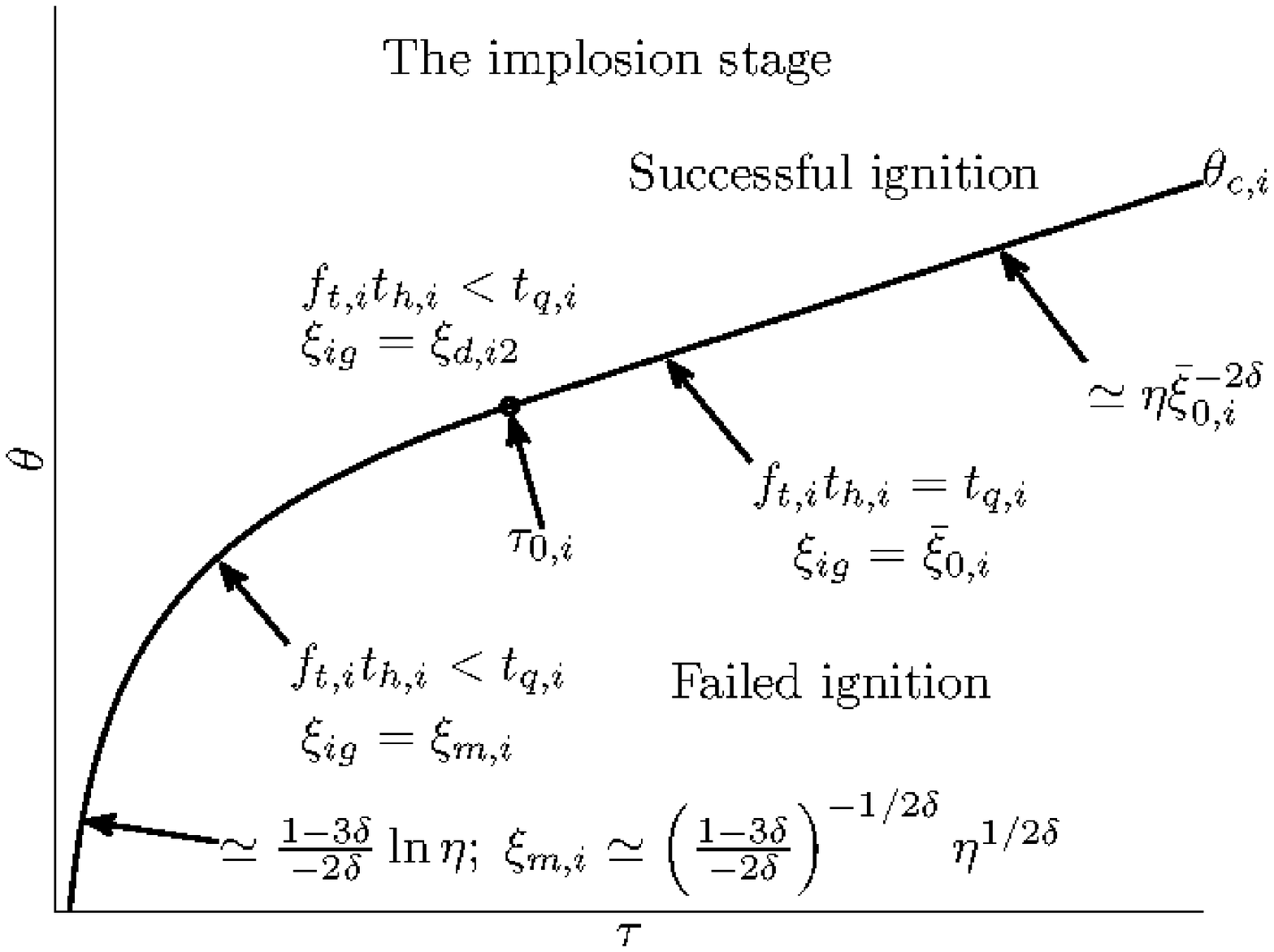}{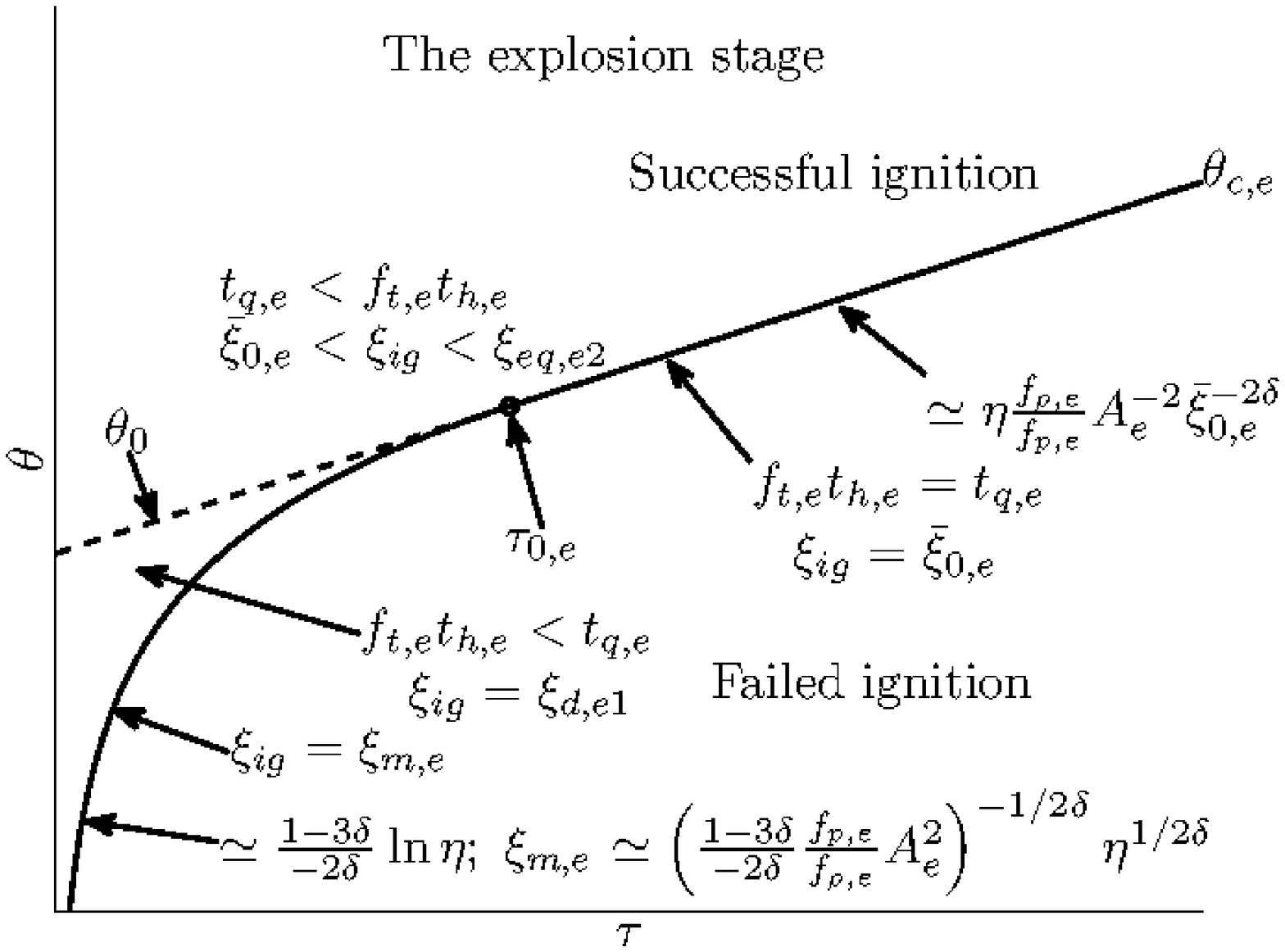} \caption{Qualitative description of the flow behavior at different parts of the $\left(\tau,\theta\right)$-plane. We expect ignition near $\xi=\xi_{\textrm{ig}}$.
\label{fig:estimates_imp}}
\end{figure}

A qualitative description of the flow behavior at different parts of the $\left(\tau,\theta\right)$-plane is given in the left panel of Figure~\ref{fig:estimates_imp}. Successful ignition is predicted at $\xi=\xi_{\textrm{ig}}=\xi_{d,i2}$ for $\theta>\theta_{c,i}$. For large values of $\tau$ ($\eta$), we obtain $\theta_{c,i}\simeq\eta\bar{\xi}_{0,i}^{-2\delta}$, and critical ignition takes place near $R=\bar{\xi}_{0,i}R_{\textrm{CJ}}$. For small values of $\tau$, critical ignition takes place near $\xi_{m,i}\propto\tau^{-1/2|\delta|}\gg1$ and $\theta_{c,i}\propto\ln\tau$. In this case ignition is expected at $R\sim \xi_{m,i} R_{\textrm{CJ}}\sim \tau^{3/2}D_{\textrm{CJ}}/\kappa$ with $Q_{h,i}/Q\sim\tau\ll1$.

\subsection{The Explosion Stage}
\label{sec:explosion_stage}

As explained in Section~\ref{sec:estimates}, the exploding shock may propagate into burnt material, even if ignition is not achieved during implosion. Thus, for the explosion case we require Equation~\eqref{eq:hydro condition2} to be satisfied at a radius where the reflected shock propagates into unburnt material, i.e., into material for which $t_{q}$ in the upstream is larger than $t_{h,e}$. For simplicity, let us first assume that $t_q$ in the upstream of the shock is similar to $t_q$ in the downstream,($t_{q,e}$). This is not a bad assumption since the reflected shock has a finite Mach number, so that the pre- and post-shock density and temperature are typically not very different. We later comment on modifications of our conclusions due to the inequality of the pre- and post-shock value of $t_q$.

\begin{figure}
\epsscale{1} \plottwo{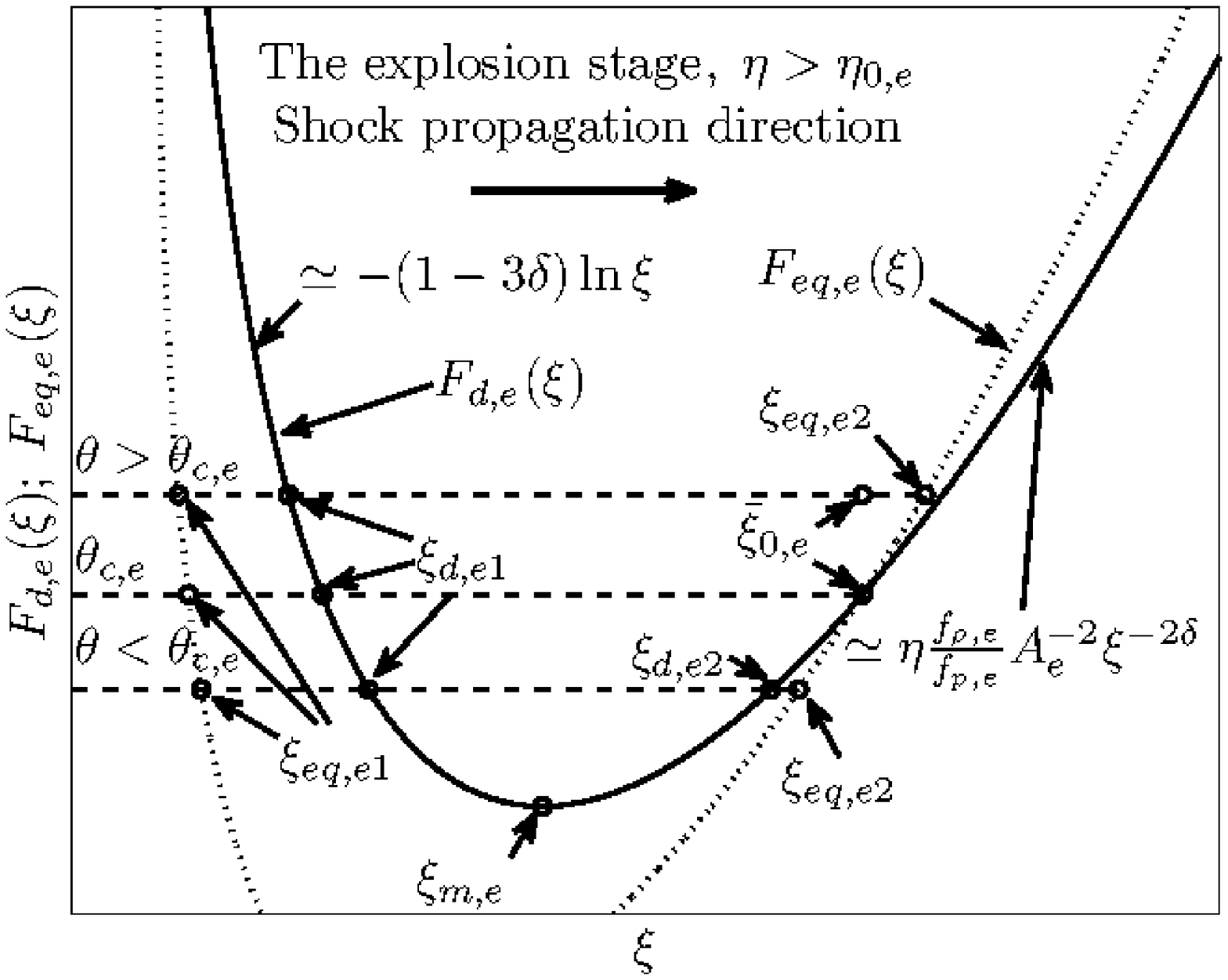}{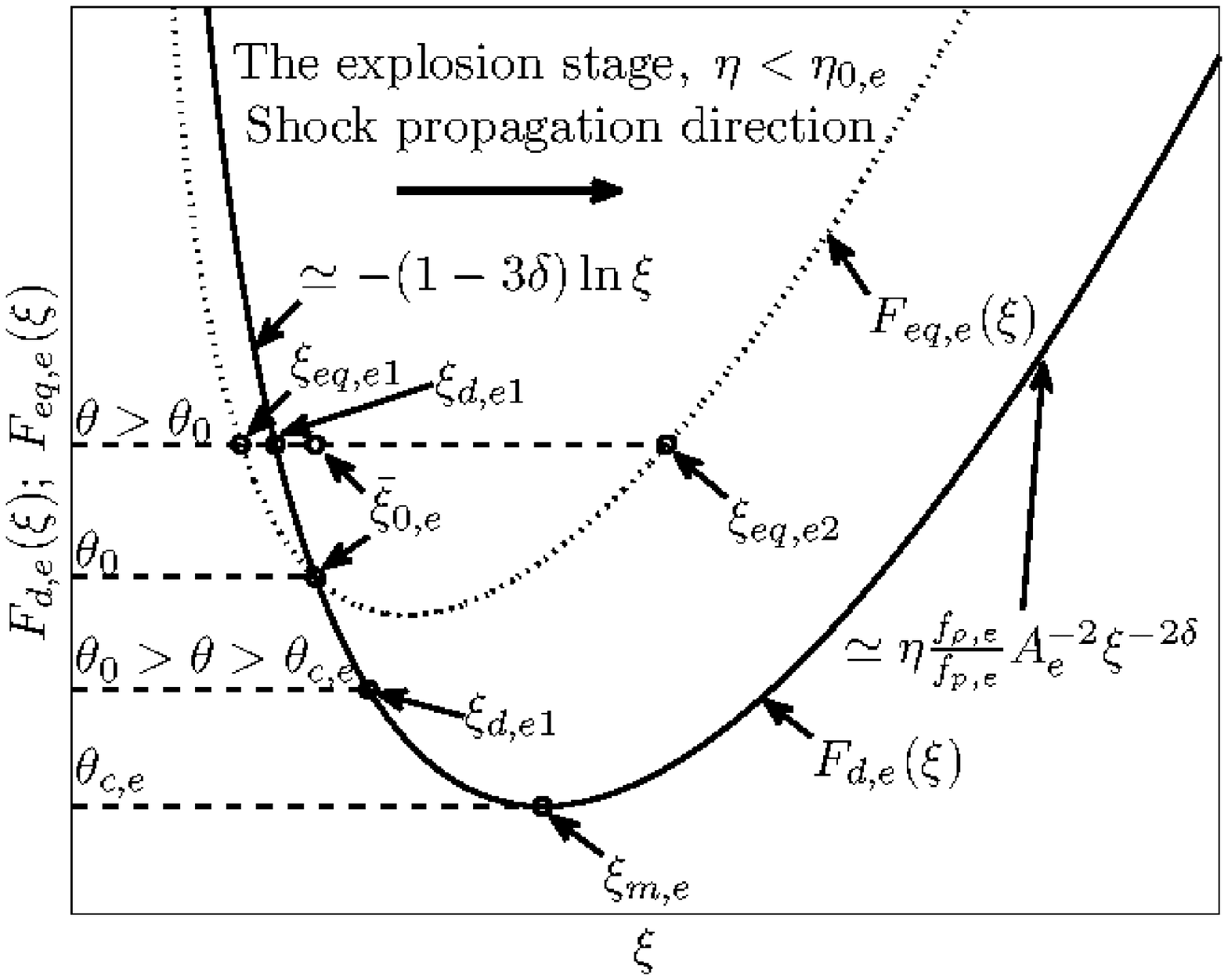} \caption{Same as Figure~\ref{fig:gxi_imp1}, but for the explosion stage.
\label{fig:gxi_exp1}}
\end{figure}

Let us first consider the $\eta<\eta_{0,e}$ case, which is illustrated on the right panel of Figure~\ref{fig:gxi_exp1} and for which $\bar{\xi}_{0,e}<\xi_{m,e}$ and $\theta_{c,e}= F_{d,e}(\xi_{m,e})$. For this case it is useful to define $\theta_{0}\equiv F_{d,e}(\bar{\xi}_{0,e})=F_{\textrm{eq},e}(\bar{\xi}_{0,e})$. For $\theta>\theta_{0}$ we expect that ignition would not be achieved at $\xi=\bar{\xi}_{0,e}$, although Equation~\eqref{eq:hydro condition2} is satisfied there, since the reflected shock is likely to propagate into burnt material at $\xi=\bar{\xi}_{0,e}$. Instead, we expect ignition only at $\xi=\xi_{\textrm{eq},e2}$. For $\theta_{0}>\theta>\theta_{c,e}$, on the other hand, we expect ignition at  $\xi_{d,e1}$.

The $\eta>\eta_{0,e}$ case, for which $\bar{\xi}_{0,e}>\xi_{m,i}$ and $\theta_{c,e}=F_{d,e}(\bar{\xi}_{0,e})=F_{\textrm{eq},e}(\bar{\xi}_{0,e})$, is illustrated in the left panel of  Figure~\ref{fig:gxi_exp1}. Arguments similar to those applied for the $\theta>\theta_{0}$ and $\eta<\eta_{0,e}$ case imply that for $\theta>\theta_{c,e}$ ignition is likely to be achieved near $\xi=\xi_{\textrm{eq},e2}$. For this case all ignitions are obtained with $t_{q,e}\sim f_{t,e}t_{h,e}$.

A comment is in place here regarding the assumed equality of $t_q$ in the upstream of the shock and $t_{q,e}$. Since the density and the temperature in the upstream of the shock are lower than the downstream values, $t_q$ in the upstream of the shock is larger than $t_{q,e}$. This implies that the shock propagates onto unburnt material for some $\bar{\xi}_{0,e}<\xi<\xi_{\textrm{eq},e2}$, and therefore ignition is likely obtained in that range. As noted above, since the Mach number of the reflected shock is finite, the post and pre-shock conditions are typically not very different, so that ignition is likely to be achieved near $\xi_{\textrm{eq},e2}$. Furthermore, since $\bar{\xi}_{0,e}=\xi_{\textrm{eq},e2}$ for $\theta_{c,e}$, our estimates for the critical value of $\theta$ required for ignition are not affected by the inequality of $t_q$ in the upstream and $t_{q,e}$.

A qualitative description of the flow behavior at different parts of the $\left(\tau,\theta\right)$-plane is given in the right panel of Figure~\ref{fig:estimates_imp}. In this case ignitions are obtained with $t_{q,e}\sim f_{t,e}t_{h,e}$ ($f_{Q,e}Q_{h,e}/Q\ge1$) for $\eta>\eta_{0,e}$ and for $\{\theta>\theta_{0},\eta<\eta_{0,e}\}$. Otherwise, ignitions are obtained with $f_{t,e}t_{h,e}\le t_{q,e}$. Above the critical curve for $\eta>\eta_{0,e}$, and above $\theta_{0}$ for $\eta<\eta_{0,e}$, we expect ignition
near $\xi_{\textrm{eq},e2}$. In the region between $\theta=\theta_{0}$ and the critical curve for $\eta<\eta_{0,e}$ we expect ignition at $\xi=\xi_{d,e1}$. For $\theta\rightarrow\infty$ we have $\xi_{\textrm{eq},e2}\simeq(f_{p,e}A_{e}^{2}\theta/f_{\rho,e}\eta)^{-1/2\delta}\rightarrow\infty$, and ignition is obtained for $R\gg R_{\textrm{CJ}}$. For large values of $\tau$ ($\eta$), we obtain $\theta_{c,e}\simeq\eta f_{\rho,e}f_{p,e}^{-1}A_{e}^{-2}\bar{\xi}_{0,e}^{-2\delta}$, and critical ignition takes place near $R=\bar{\xi}_{0,e}R_{\textrm{CJ}}$. For small values of $\tau$, critical ignition takes place near $\xi_{m,e}\propto\tau^{-1/2|\delta|}$ and, similarly to the implosion case, $\theta_{c,e}\propto\ln\tau$ and ignition is expected at $R\sim \xi_{m,i} R_{\textrm{CJ}}\sim \tau^{3/2}D_{\textrm{CJ}}/\kappa$ with $Q_{h,i}/Q\sim\tau\ll1$.

\section{The numerical model}
\label{sec:numerical model}

We used the one-dimensional, Lagrangian version of the VULCAN code \citep[for details, see][]{Livne1993IMT}, with a simple extension to include reaction rates of the form of Equation~\eqref{eq:rate}. The initial conditions used are zero velocity, constant density $\rho_{0}$, and small pressure $p_{0}$ (see below) throughout the computation's region $r<L_{p}$. The initial mesh spacing was uniform with $N$ cells, and we did not use any remapping or AMR technics. The outermost node acted as a piston with constant inward velocity, $v_{p}$, throughout the simulation. The initial pressure was chosen such that the initial sound speed was $\simeq0.1v_{p}$. This ensures that initially the ratio between the upstream pressure and the downstream pressure of the imploding shock wave is $\simeq0.01$. A further reduction of the initial pressure did not change significantly any of our results, but increased the computing time due to the Courant condition. We also suppressed to zero the reaction rate for $p/\rho<1.001p_{0}/\rho_{0}$ in order to ensure that the material is completely unburnt until it is shock heated.

The values of $L_{p}$, $v_{p}$, and $\rho_{0}$, which determine the dimensional scales of the problem, were chosen arbitrarily and were held fixed for all of our simulations. Given the values of $\nu$, $n$, $m$, and $\gamma$, we used the following consideration to set the values of $\kappa$, $p_{A}$, and $Q$ in order to calculate the flow for any choice of $\left(\tau,\theta\right)$. The velocity of the shock wave as it emerges from the piston is
\begin{eqnarray}\label{eq:Num1}
\dot{R}(L_{p})\simeq\frac{\gamma+1}{2}v_{p}\equiv f_{v}v_{p}.
\end{eqnarray}
Hence
\begin{eqnarray}\label{eq:Num2}
f_{v}v_{p}=-D_{\textrm{CJ}}\left(\frac{L_{p}}{R_{\textrm{CJ}}}\right)^{\delta},
\end{eqnarray}
and we have
\begin{eqnarray}\label{eq:Num3}
D_{\textrm{CJ}}=-f_{v}v_{p}K^{-\delta},
\end{eqnarray}
where $R_{\textrm{CJ}}=L_{p}/K$ (we discuss below how to choose the value of $K$). The CJ velocity sets the value of $Q$, and Equation~\eqref{eq:tau def} sets the values of $p_{A}$. The value of $\kappa$ is given now by Equation~\eqref{eq:theta}.

In order to determine if a successful ignition took place, we investigate the behavior of the outgoing waves. Our criterion for a successful ignition is that a detonation wave with $w_{\textrm{Num}}/w_{\textrm{ZND}}<w_{\textrm{th}}$ (see Section~\ref{sec:numerical}) has propagated over at least 100 reaction zone lengths. Since these waves ultimately hit the piston, we have to make sure that the outgoing waves either satisfy the criterion for a successful ignition or that $w_{\textrm{Num}}/w_{\textrm{ZND}}$ diverges with time, before the piston is reach by the wave. Increasing the available distance for the outgoing wave (where at some point a decision regarding whether a successful ignition took place can be made) can be done by increasing the value of $K$. However, since we want to solve for the reaction zone behind a possible detonation wave with a given resolution, increasing $K$ also increases $N$. The value of $K$ was chosen to be the minimal value for which a decision regarding whether a successful ignition took place can be made, and the value of $N$ was chosen such that the reaction zone was fully resolved (we always checked for convergence of our results, and in general we found that $\simeq100$ cells are sufficient for a converged calculation of the reaction zone). With the method presented here we were able to calculate all cases of interest with $N<5\times10^{4}$.


\bibliographystyle{hapj}
\bibliography{ms}


\end{document}

%% file: Definitions.tex
\newcommand{\DDir}{\relax{D\kern-.7em{/}}}

\newcommand{\be}{\begin{equation}}
\newcommand{\ee}{\end{equation}}
\newcommand{\bea}{\begin{equation*}}
\newcommand{\eea}{\end{equation*}}

\newcommand{\nin}{\relax{\in\kern-.8em{/}}}

